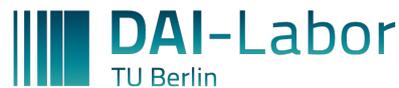

# Master Thesis

# Reinforcement Learning control strategies for Electric Vehicles and Renewable energy sources Virtual Power Plants

Francesco Maldonato

| | |
|---|---|
| Examiner: | Prof. Dr. Dr. h.c. Sahin Albayrak |
| | Prof. Dr. Odej Kao |
| Supervisor: | M.Sc Izgh Hadachi |
| Student N.: | 474370 |
| Date: | 25. October 2022 |

ICT Innovation

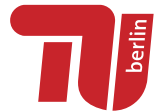

Technische Universität Berlin
Faklutät IV - Elektrotechnik und Informatik
Institut für Telekommunikationssysteme, Fachgebiet AOT

EIT Digital - Autonomous Systems (AUS)

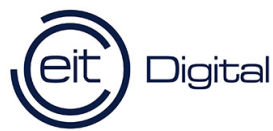

# Eidesstattliche Versicherung (affidavit)

Hiermit erkläre ich, dass ich die vorliegende Arbeit selbstständig und eigenhändig sowie ohne unerlaubte fremde Hilfe und ausschließlich unter Verwendung der aufgeführten Quellen und Hilfsmittel angefertigt habe.

I hereby declare that I have written the present work independently and by my own hand without any unauthorized help and exclusively using the resources listed.

Berlin, 25 October 2022

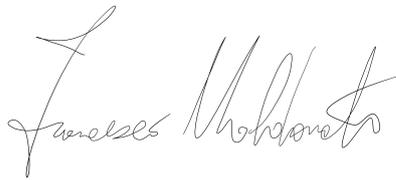

---

Francesco Maldonato

i

# Ringraziamenti

Molte persone mi sono state vicine durante questi anni di studio e, in un modo o nell'altro, mi hanno aiutato in questo percorso.

A mamma e papà, al loro costante sostegno ed ai loro insegnamenti senza i quali oggi non sarei ciò che sono. Senza di voi, tutto questo non sarebbe stato possibile. Ai miei fratelli Marco e Andrea, che mi hanno insegnato a condividere, dalla felicità ai dolori. A mio zio Alessandro, che mi dà speranza al non invecchiare mai.

A Simone, Gianluca, Alessandro, Federico e Isidoro; i miei amici di sempre da Marsala con cui sono cresciuto e con cui mi sento sempre a casa.
A Lurenz, folle partner di avventure a Shanghai, a Dario, mitico compagno dell'epopea in Asia, ed ai grandi amici di Torino: Diba, Ciccio Piazza, Angiluzzo, Rebecca, Paolino e Anna. Siete stati come una famiglia, prendendoci cura a vicenda e condividendo gli anni più divertenti alla Madhouse.
Ad Otto, che è stato come un fratello durante l'anno in Francia a Nizza.
A Zaida, per tutto il supporto emotivo che mi regala in questo viaggio.

A mio nonno Franco, che so che sarebbe tanto orgoglioso.

# Acknowledgements

I would like to thank all the friends that had been close to me during these years of studies abroad. To the EIT Nice gang in particular, a group of special people with whom I spent a great year, even with the Covid lockdown.

I am grateful to the Technische Universität Berlin department of DAI-Labor that allowed me to develop this thesis and it proved to be the perfect environment for such research.

I thank the OpenAI and Stable Baselines3 community for providing open-source reinforcement learning tools, and Weights and Biases for providing a free academic license to track the research progress.

A special thank you goes to my thesis tutor Izgh Hadachi, who gave me great help, inspiration, and thrust to pursue this thesis, and to my supervisor Nadim El Sayed, who guided me along the work done at DAI-Labor and taught me a lot.

# Zusammenfassung

Die steigende Nachfrage nach direkter elektrischer Energie im Stromnetz hängt auch mit der zunehmenden Nutzung von Elektro-Fahrzeugen in den Städten zusammen, die Fahrzeuge mit Verbrennungsmotor schließlich vollständig ersetzen werden. Allerdings wird diese große Menge an benötigter Energie, die in den Batterien der Elektro-Fahrzeuge gespeichert ist, nicht immer genutzt und kann ein eigenes virtuelles Kraftwerk darstellen. Wenn das Stromnetz Zugang zu den Batterien hat, kann es diese als Backup-Stromquelle bei Überlast oder als zusätzlichen Speicher bei Unterlast nutzen und so die Wahrscheinlichkeit von Stromausfällen effektiv verringern.

Bidirektionale Elektro-Fahrzeuge, die mit Batterien ausgestattet und an das Stromnetz angeschlossen sind, können daher Energie je nach Bedarf laden oder entladen und so eine intelligente Verlagerung der Energie dorthin bewirken, wo sie benötigt wird. Elektro-Fahrzeuge, die als mobile Speicher eingesetzt werden, können die Ausfallsicherheit und das Gleichgewicht zwischen Angebot und Nachfrage für bestimmte Verbraucher verbessern, in vielen Fällen als Teil eines Microgrid. Je nach Richtung des Energietransfers können Elektro-Fahrzeuge Haushalte durch Laden von Fahrzeug zu Hause mit Reservestrom versorgen oder ungenutzten Strom aus erneuerbaren Energien durch Laden von Fahrzeug zu Fahrzeug speichern. Diese Lösungen können erneuerbare Energiequellen wie photovoltaische Solaranlagen und Windturbinen, die im Laufe der Zeit schwanken, ergänzen, den Eigenverbrauch und die Autarkierate erhöhen und dem Ziel von Netto-Null-Energie-Gebäuden näher kommen.

Das Konzept der verteilten Energieressourcen wird immer präsenter und erfordert neue Lösungen für die Integration mehrerer komplementärer Ressourcen mit zeitlich variablem Angebot. Die Entwicklung dieser Ideen geht mit dem Wachstum neuer KI-Techniken einher, die potenziell den Kern solcher Systeme bilden werden. Techniken des maschinellen Lernens können die Umgebung des Energienetzes so flexibel modellieren, dass eine ständige Optimierung möglich ist. Dieses faszinierende Arbeitsprinzip führt das umfassendere Konzept eines vernetzten, gemeinsam genutzten, dezentralen Energienetzes ein, das perfekt in Systeme mit verteilten erneuerbaren Ressourcen integriert werden kann.

Diese Forschung über Reinforcement Learning Kontrollstrategien für Elektrofahrzeuge und erneuerbare Energiequellen konzentriert sich auf die Bereitstellung von Lösungen für solche Modelle zur Optimierung der Energieversorgung.

# Abstract

The increasing demand for direct electric energy in the grid is also tied to the increase of Electric Vehicle (EV) usage in the cities, which eventually will totally substitute combustion engine Vehicles. Nevertheless, this high amount of energy required, which is stored in the EV batteries, is not always used and it can constitute a virtual power plant on its own. Having access to the EV batteries, the power grid can potentially exploit them as a backup power source in case of overload, or as an additional storage device in case of being under-loaded, effectively decreasing the probability of black-outs.

Bidirectional EVs equipped with batteries connected to the grid can therefore charge or discharge energy depending on public needs, producing a smart shift of energy where and when needed. This concept has become feasible thanks to new bi-directional power infrastructures with high power rates. EVs employed as mobile storage devices can add resilience and supply/demand balance benefits to specific loads, in many cases as part of a Microgrid (MG). Depending on the direction of the energy transfer, EVs can provide backup power to households through vehicle-to-house (V2H) charging, or storing unused renewable power through renewable-to-vehicle (RE2V) charging. V2H and RE2V solutions can complement renewable power sources like solar photovoltaic (PV) panels and wind turbines (WT), which fluctuate over time, increasing the self-consumption and autarky rates and getting closer to the Net-Zero-Energy Buildings (NZEBs) goal.

The concept of distributed energy resources (DERs) is becoming more and more present and requires new solutions for the integration of multiple complementary resources with variable supply over time. The development of these ideas is coupled with the growth of new AI techniques that will potentially be the managing core of such systems. Machine learning techniques can model the energy grid environment in such a flexible way that constant optimization is possible. This fascinating working principle introduces the wider concept of an interconnected, shared, decentralized grid of energy.

This research on Reinforcement Learning control strategies for Electric Vehicles and Renewable energy sources Virtual Power Plants focuses on providing solutions for such energy supply optimization models.

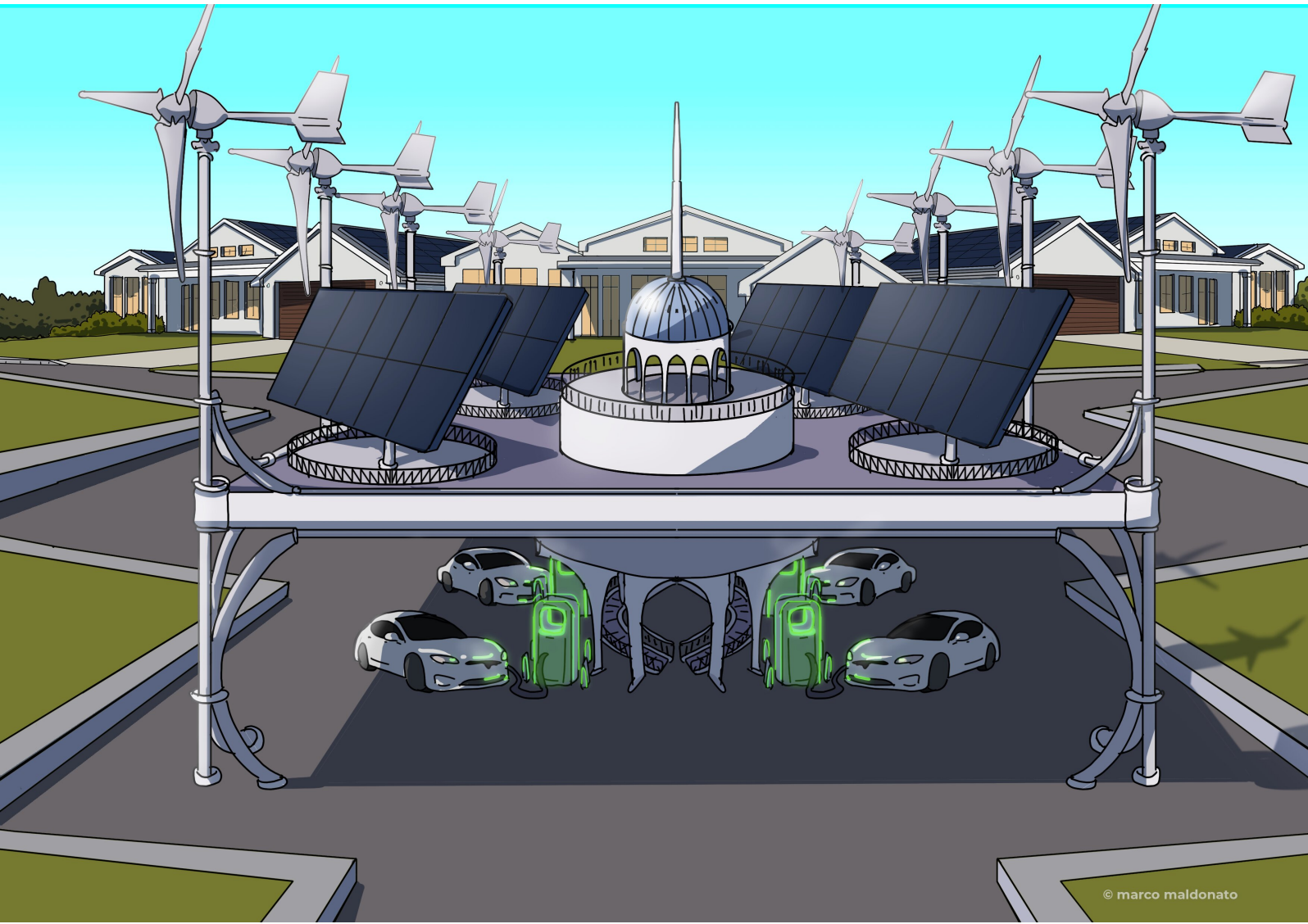

v

# Motivation and purpose

A fundamental step in the Innovation process is understanding future trends, needs, and behaviours of society. The climate change crisis is making some of them very clear by setting new goals that humanity must achieve to survive on the planet, i.e. reaching carbon neutrality by 2050. The main issue to solve in order to achieve the zero-carbon emissions goal is related to energy production and management: the old, centralized, and static power plant system to distribute energy is outdated and inefficient. A new, decentralized, smart, and flexible power plant must be designed to satisfy the increasing demand for energy to power future cities.

It is my great concern to help human society reduce its impact on the environment, while still developing its technological horizons. Considering the increasingly high number of inhabitants in today's metropolis, this can only be possible through massive cooperative innovation and maximum efficiency. In the power grid case, new levels of efficiency can be reached by rethinking the system with new AI approaches.

This project has been driven by a fascination for Artificial Intelligence, and a deep interest in the implementation of renewable energy smart grids. Particular attention is paid to the sustainability of Renewable energy sources, which are a key factor to understand the limit of natural resources and a priority to fight climate change.

# Abbreviations and Symbols

|                | | |
|---|---|---|
| | ML | Machine Learning |
| | RL | Reinforcement Learning |
| | VPP | Virtual Power Plant |
| | EV | Electric Vehicle |
| | EVSE | Electric Vehicle Supply Equipment |
| | V2B | Vehicle-to-Building |
| | V2H | Vehicle-to-Home |
| | V2G | Vehicle-to-Grid |
| | MG | Micro-Grid |
| | AG | Aggregator |
| | CS | Charging Station |
| | ESS | Energy Storage System |
| | HJB | Hamilton-Jacobi-Bellman |
| | MDP | Markov decision process |
| | MRP | Markov reward process |
| *Abbreviations* | POMDP | Partially observable Markov decision process |
| | NN | Neural Network |
| | RNN | Recurrent Neural Network |
| | DNN | Dense Neural Network |
| | DERs | Distributed Energy Resources |
| | PV | Photovoltaic |
| | WT | Wind Turbine |
| | FC | Fuel Cell |
| | SoC | State of Charge |
| | RE2V | Renewable-to-vehicle(EV) |
| | RE2B | Renewable-to-building |
| | RE2H | Renewable-to-home |
| | FLC | Fuzzy Logic Controllers |
| | LSTM | Long Short Term Memory |
| | NZEB | Net Zero Energy Buildings |
| | V2X | Vehicle(EV)-to-any |

# Contents

# List of Figures

# Chapter 1

# Introduction

This chapter describes the context of the problem, the goals of this project, the specific problem that this research addresses, and its main contributions. It finally outlines the structure of the thesis.

## 1.1 Context

### 1.1.1 Power demand and Electric Vehicles adoption

The growing integration plans of renewable energy sources and EVs in the urban power grid by 2035 will have a significant impact on network stability. McKinsey's study "Global Energy Perspective 2021" [1] estimated that by 2035 there will be an increase of approximately 23.8% (1.7% of terawatt-hours per year) of the total power demand in Europe, with an enormous growth in the transport sector due to the electrification of transports and the massive adoption of Electric Vehicles as shown in the graph of the study reported in Figure 1.1. In fact, [2] in 2035 and beyond in Europe, (2040 in the rest of the world), 100% of new vehicles sold are expected to be fully electric. The study of Transport & Environment [3] stated that the COVID-19 pandemic seems to have had a positive impact on EV sales. Pre-pandemic, early research estimated that 33 Million EVs would roam around Europe by 2030. Now, Pandemic-adjusted research shows that we're looking at 40 Million electric vehicles driving around the continent by 2030. Figure 1.2 shows the expected trend in the next ten years. EV charging stations' rated powers go usually between 3.7 and 11 kW, which is considerably higher than the average household power demand. The resulting superposition of all the EVs demand connected to the grid in an urban area can be a threat to the stability of the grid itself. This quick change in the power demand comports abrupt peak load demands, assuming an uncontrolled charging of the EVs.

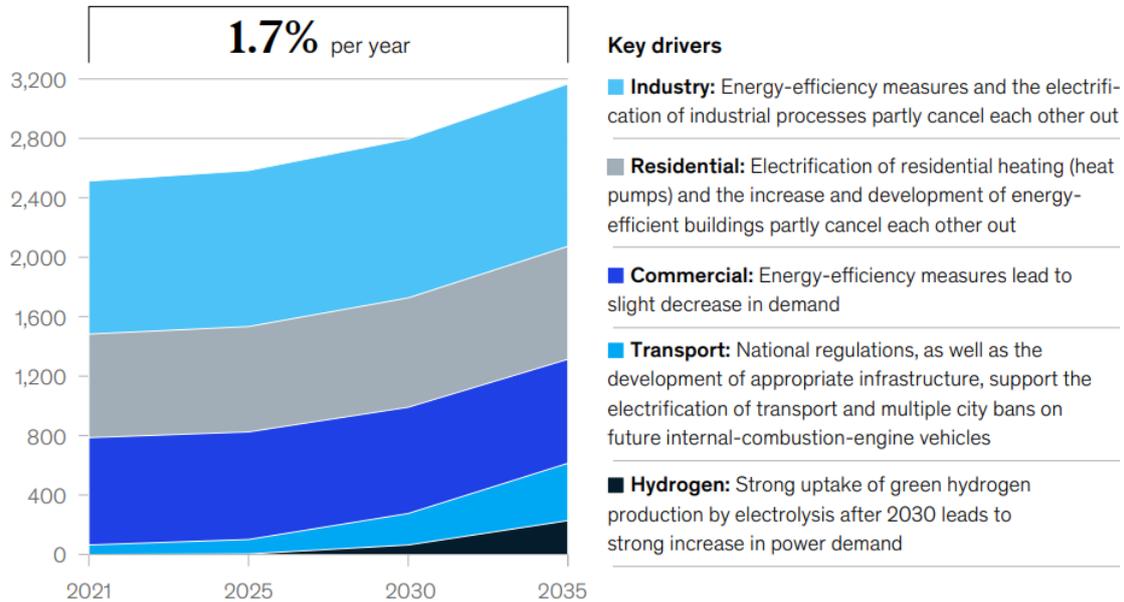

Figure 1.1: European forecasted power demand. Source: "Global Energy Perspective 2021", January 2021 McKinsey.com [1]

However, adopting bidirectional electric vehicles (EV) as mobile storage devices can add resilience and supply/demand balance benefits to specific loads. EVs can receive energy (charge) from the smart charging stations (CSs), and also push it back providing power to an external load (discharge). In this case, the electric vehicle supply equipment (EVSE) constitutes a power source of its own, or a Virtual Power Plant (VPP).

### 1.1.2 Electric Vehicles integration benefits

Bidirectional vehicles [4] can be employed as part of a Micro-Grid in specific energy transfer operations. The most popular ones are Vehicle-to-Building (V2B) and Vehicle-to-Grid (V2G), where the energy transfer happens from the EV to an external building or the public grid to provide load support. These solutions are usually coupled with renewable power sources like solar photovoltaic (PV) arrays, wind turbines (WT), and other distributed energy resources (DERs) that are not always available and fluctuate over time. In these applications, EVs can potentially store all the excess energy produced by renewable sources that otherwise would be exported to the grid causing under-loading. In any case, this energy would be just lost if unused. Moreover, EVs are light and they can easily move where their support is needed, e.g. in areas with planned outages or where they can serve as an emergency backup power source. The use-cases scenario of such solutions is represented in Figure 1.3.

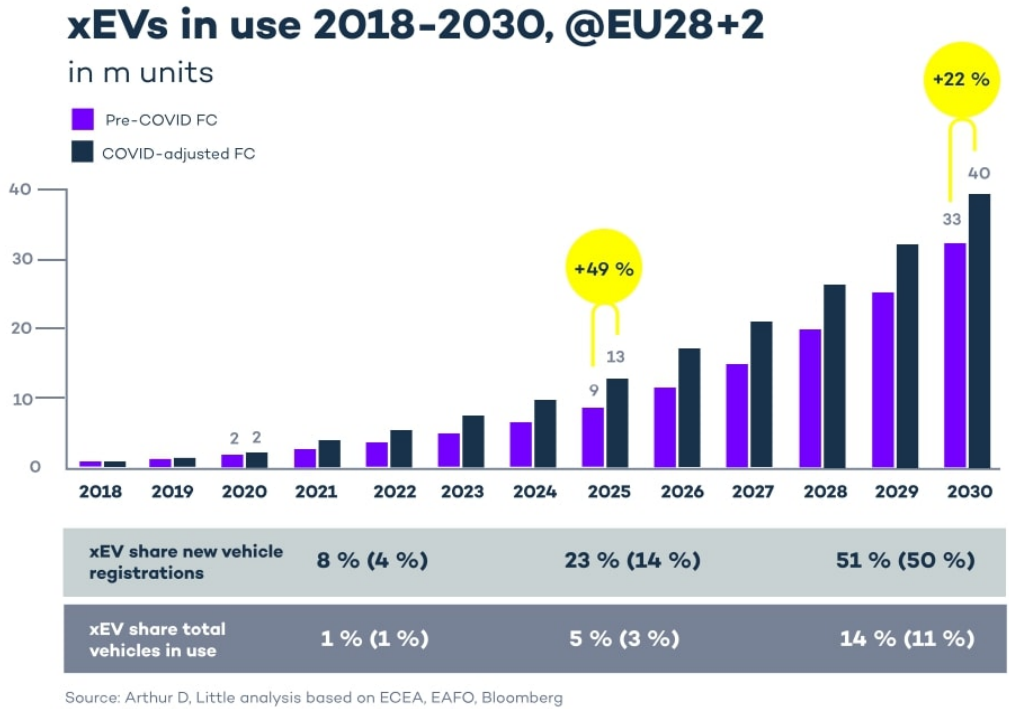

Figure 1.2: EVs usage statistics in the EU. Source: Arthur D. Little analysis based on ECEA, EAFO, Bloomberg [2]

Integrating EVs as active players in the grid is a powerful resource for the resilience and supply/demand balance of decentralized grids. In particular, EV batteries help the decarbonization of emergency backup resources, which are usually based on fossil fuels. Currently, integrating EVs as storage devices essentially helps reduce carbon emissions by maximizing the consumption of local and renewable generated power. In other words, they can increase the autarky and self-consumption rates of energy systems. In general, the decentralization of energy resources is fundamental in the urban power grid. In fact, the energy ecosystem is moving towards a more decentralized generation and distribution design, where distributed energy resources

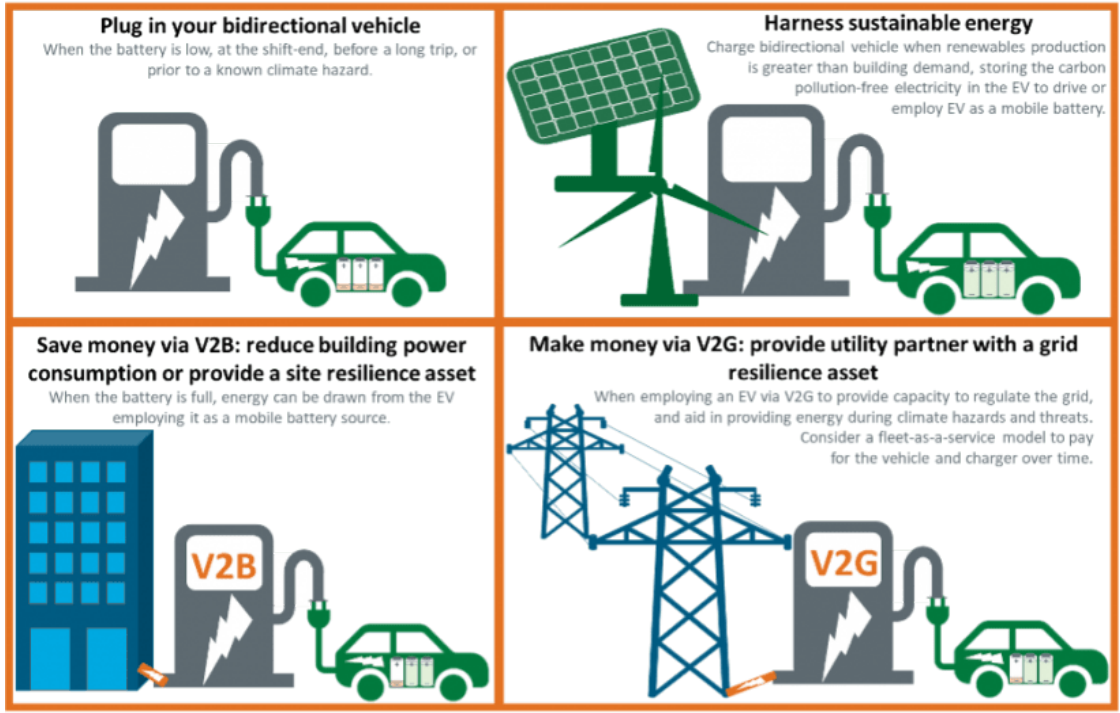

Figure 1.3: Bidirectional EVs charging operations use cases. Source: Federal Energy Management Program » Bidirectional Charging and Electric Vehicles for Mobile Storage [4]

(DERs) operate independently from the power providers and even sell excess energy back to the main grid. This behaviour is encouraged by the regulating agencies due to environmental, social, and economic benefits. A certain number of DERs can choose to bundle together forming a small electrical grid also known as a microgrid (MG) or a virtual power plant (VPP). A certain number of EVs connected to the grid constitutes a VPP itself, and the extent of how this virtual power plant is going to behave is still a subject of study.

## 1.2  Research content and scope of the thesis

As a result of the analysis of the previous section, EVs can provide significant support to the power station in several ways, especially if coupled with renewable energy sources. This can be possible by using EVs as decentralized storage devices from Renewable energy sources, which do not have constant output and availability.

[5] The purpose of using EVs is that the quick action of their batteries can respond to frequency deviations much faster than the traditional generators and hence stabilize the grid. Automation techniques can ensure this balance by switching EV batteries connected to the grid in charging/discharging mode. In particular, a Reinforcement Learning environment could be able to manage EVs' mode and can be trained to support the grid as much as possible.

**Research Goal**

The main scope of this thesis is to model an approach to control a distributed system of production, storage, and self-supply electric resources, also known as VPP. The formulated problem will have the priority of minimizing supply/demand imbalances and also ensuring a certain power left on the EVs. Specifically, the research will be focused on the integration of EVs in the grid to stabilize the load by performing the following main objectives:

- Peak shaving: pushing electric power (discharging) to the grid when demand is high.

- Valley filling: drawing electric power (charging) when the grid is under-loaded.

- Net-zero resultant load: balancing out the supply/demand load in real-time.

The EVs Virtual Power plant will therefore have the main goal of flattening the load over time to avoid imbalances and prevent the grid from power shortages. This integration will help maintain the resultant load to zero as long as possible. The second main objective of the VPP is to build a Micro-Grid system that is auto-sufficient as much as possible, so that it does not require buying energy from the public grid. Finally, the energy left on the EVs during the simulation must be sufficient to be used after leaving the charging station.

**Research question focus and contributions**

Now that the context, problem, and goal of the research are described, the most important clue to define is how the goals will be achieved, i.e. what are the solutions to be explored. Grid stability operations are actually a set of problems already explored through control theory. Frequency stabilization and over/under-loading operations are procedures that are automatically executed by the modern grid infrastructure. However, these procedures rely mainly on fossil fuels and suppose a constant availability of energy resources ready to be used.

Since the nature of renewable energy sources is fluctuant, as well as the EVs availability at the charging stations, the solution must consider the aleatory availability of resources. Therefore, a suitable solution to regulate the Virtual Power Plant could rely on Machine Learning. Such a solution can apply since both renewable energy sources and EVs availability are aleatory, but they still follow a cyclic pattern throughout the simulation that can be detected. Specifically, among the Machine Learning (ML) paradigms available, the most suited for this type of problem is Reinforcement Learning (RL), since the cyclic pattern of the variables requires an action of control theory type, aimed at the optimization of some features. Several types of RL algorithms can be applied to this optimization problem that aims to flatten the resultant load. Nevertheless, the RL solution for this problem should answer the following questions:

- What are the bottlenecks of a self-sufficient VPP relying on renewable sources and EVs?

- How many EV charging events must occur to guarantee self-sufficiency?

### 1.2.1 Simulation Scenario

The hypothetical scenario of the formulated problem to be optimized is composed of 3 main elements acting in the simulation:

- The renewable production system: a fluctuating energy production system composed of 40 Solar panels units having 400 W of power capacity each, and 8 urban installable wind turbines with 1500 W capacity each.

- The household consumers: 4 Households connected to the grid that will constantly require power.

- The Bidirectional Electric vehicle supply equipment (EVSE): A set of 4 EV charging stations directly connected to the MG, where vehicles can connect over time, and are able to charge or discharge.

The visualization of the described scenario is available in Figure 1.4.

### 1.2.2 Code development

The code for the execution of the VPP is entirely written in Python. The VPP environment class and relative functions are all defined in the $VPP\_environment.py$ script. The rest of the repository files are VPP simulator, agent trainer, or hyper-parameters tuner Jupyter notebooks. The source code repository and more details are presented in the Appendix.

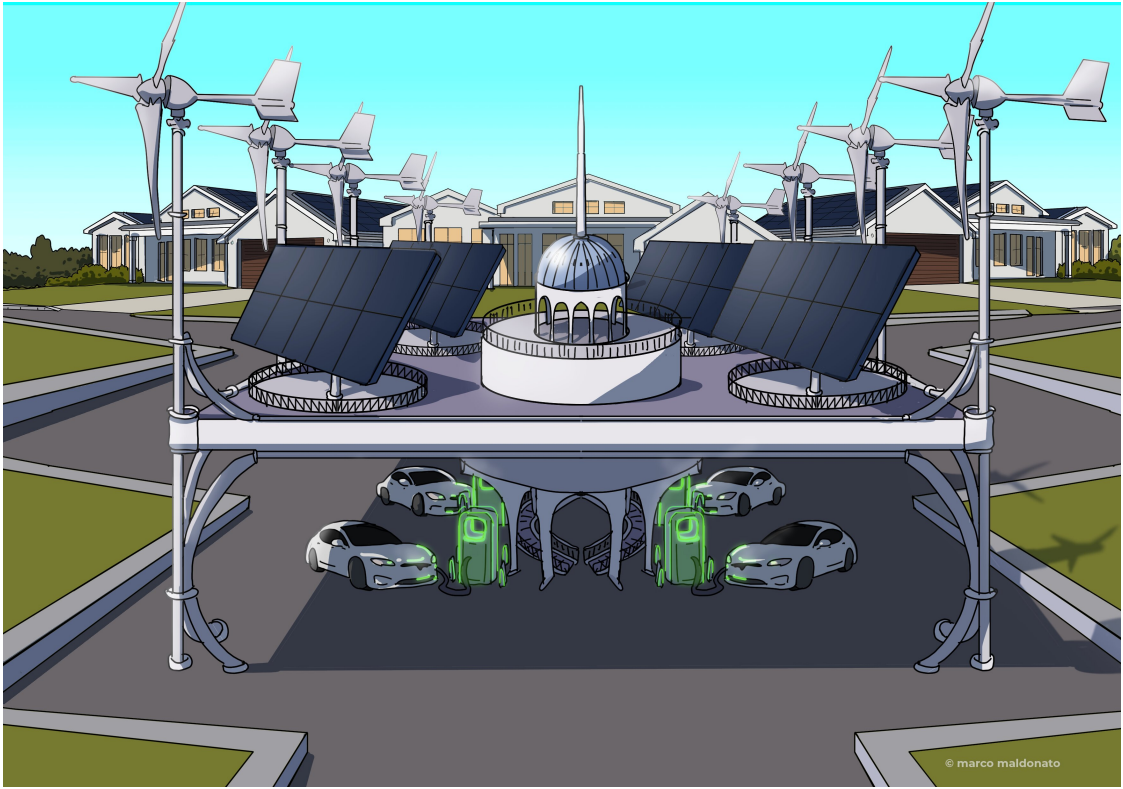

Figure 1.4: EV Virtual Power Plant scenario

### 1.2.3 Research outline

The contents of the following chapters can be summarized as follow:

- Chapter 2 introduces the required technical background needed for this research. An overview of the state of the art of applications of EVs supporting the grid is given, as well as the relevant Reinforcement Learning theory used. In particular, the RL framework used, the concept of policy-gradient based models, and the main concepts necessary to understand how the used algorithms work.

- Chapter 3 focuses on the analysis of the research. Specifically, the analysis of the starting point of work and the steps to complete to have a working environment. It continues with the description of the data-sets used and the pre-processing applied.

- Chapter 4 describes the environment shape and the definition process that took place during the research. it continues with the description of the RL agent structure and features developed. It ends with a description of the final solution proposed for the Virtual Power Plant problem.

- Chapter 5 dives into the main task of the thesis: the implementation of the Reinforcement Learning algorithms and the outcomes. It describes the training algorithms strategy and it develops a comparison between the performances of the algorithms. It concludes with the proposal of the best-suited algorithm for the solution to be implemented.

- Chapter 6 describes and explains the results obtained by the research question experiments performed on the best-suited trained model. It aims to present some benchmarks that can support the algorithm choices and evaluate the general performance of the Virtual Power Plant solution proposed in the Methodology section.

- Chapter 7 proposes some experiments to test the solution proposed for the VPP environment and to tune the Elvis configuration parameters. It continues with the analysis and discussion of the results of the evaluation of the experiments.

- Chapter 8 summarizes the results and analysis as a conclusion for such a VPP solution and introduces future work to further utilize the findings of this thesis.

## Chapter 2

# Background research

This chapter introduces the required technical background needed for this research. It begins with the available studies on EV applications on power grid support and the optimizations that have been studied in this field. Then it continues with the relevant Reinforcement Learning theory used: the description of the RL paradigm, the concept of policy-gradient based models, and the main equations necessary to understand the model algorithms used.

## 2.1 Smart electric vehicles charging literature

The Vehicle-to-grid (V2G) technology is already subject of study for many years and proved to bring a lot of benefits for power grids and the environment. The most explored are: reactive power support, active power regulation, frequency support, harmonic filtering, voltage control, spinning reserve, load balancing, and improving stability, efficiency and reliability. EVs were also studied as storage devices to minimize charging costs or to maximize revenues by selling the energy back to the grid.

This section provides an overview of all the studies on EVs employment in grid operations available in the literature with the application of **reinforcement learning** or **control theory**. However, it is important to remember that this research focuses on EVs employment in helping to balance supply and demand by valley filling and peak shaving, as it was introduced in the previous chapter.

### 2.1.1 Reinforcement learning and optimal control applied to Electric Vehicles smart charging operations

**Frequency-regulation application**

Studies are showing that the intermittency issues of Smart grids can be handled effectively with the employment of EVs. [5] By regulating the charging/discharging rates of the EVs, it is possible to provide frequency support to the grid. These regulations are controlled by the aggregators (AGs) at the charging stations (CSs) level. The research contributions on frequency fluctuations compensation of the power grid through Evs were a great starting point for this research, as this goal can be addressed as the objective of optimization techniques through charging/discharging actions. However, even if this knowledge inspired this work, the problem proposed can be solved by plain control theory rather than reinforcement learning.

**Market-oriented application**

There are several scientific papers investigating algorithms to generate revenues by charging EV batteries and selling electricity back into the grid at peak times with higher prices [6]. The techniques used are based on Reinforcement Learning, modeled with a priori or a posteriori knowledge of future grid supply/demand states. The proposed algorithms in this area of research aim to maximize profit, or in general to minimize costs.

**Model predictive control with Q-learning solutions**

There are other approaches for Electric Vehicles Integration in Micro Grids (MG) with smart charging stations that focus on several aspects. [7] Such solutions aim to dynamically control the power demand and distributed energy resources to improve the matching performance between renewable power generation and households consumption, while also trading electric energy in both day-ahead and real-time markets to reduce operational costs. In such an application, the aggregator manages a set of resources, including a photovoltaic system, energy storage system, thermostatically controllable loads, and electrical vehicles. The problem was formulated as a mixed-integer linear programming problem in which the objective is to minimize the operation and the degradation costs related to the energy storage system and the EV batteries. To mitigate the uncertainties associated with system operation, a two-level model predictive control (MPC) integrating a Q-learning reinforcement learning model is designed to address different time-scale controllers. MPC algorithm allows making decisions for the day ahead, based on predictions of uncertain parameters, whereas the Q-learning algorithm addresses real-time decisions based on real-time data.

**Smart charging stations coupled with Fuzzy logic controllers**

Also, [8] it is possible to find proposals for smart Charging Stations (CSs) coupled with Fuzzy Logic Controllers (FLCs) for charging and discharging EVs. These CS are integrated with renewable energy resources in order to support grid operations and provide eco-friendly means to manage energy. The designed FLC computes the permissible power that can be exchanged between the CS and the EVs during peak and off-peak hours. This computation is based on the node's voltage, the energy available at CS and the power to be compensated by CS. Moreover, an algorithm has been designed in order to prevent discharging of EVs beyond a specified end-of-discharge voltage. It also encourages charging and discharging at specific charging rates. Such info will provide the base for the Virtual Power Plant (VPP) environment subject of study.

**Multi-agent-system smart grid approach**

[9] Studies on VPP environments modeled using reinforcement learning on multi-agent systems (MAS) exist. In particular, Q-learning solutions to manage distributed energy resources (DER) including solar PV and wind turbine modules, diesel generators, fuel cells (FCs), and EVs [10]. In a multi-agent environment, each agent uses RL to interact with each other in a distributed manner. For implementing a de-centralized structure, there are hierarchical strategies and distributed strategies where local agents have different degrees of intelligence depending on their roles. This is a different approach from the central agent that manages all the resources in the MG with all the degrees of freedom. The multi-agent approach control structure goes in a different direction in this research since a multi-agent approach does not guarantee total control of the shared resources.

**Long-Short-Term Memory (LSTM) Reinforcement Learning for Microgrid management**

Finally, [11] LSTM Reinforcement Learning strategies for Energy Management in Microgrid implementing Electric Vehicles as storage devices are also analyzed in the literature. The approach, in this case, was to develop a reinforcement learning (RL) algorithm to smartly control the Micro-Grid's energy storage system (ESS) in real-time by considering the future reward of a charging/discharging action. The problem modeling of this solution was helpful in the analysis of this research, as well as the training strategy of the RL agent. In fact, this solution adopts a prediction model using long short-term memory (LSTM) networks to speed up the RL training stage (online optimization) and to explore the system input traces for more accurate future reward counting in the current learning process.

## 2.2 Reinforcement Learning background theory

**Reinforcement Learning as a branch of optimal control theory**

Reinforcement Learning theory is in practice a branch of optimal control, which is a concept extending from control theory, [12] used to achieve an optimal control law for a problem. Essentially, it is based on a minimization problem that aims to optimize a certain parameter. The core of optimal control theory is described in Richard Bellman's equation of Hamilton-Jacobi-Bellman (HJB):

$$\dot{V}(x,t) + \min_{u}\{\nabla V(x,t) \cdot F(x,u) + C(x,u)\} = 0, \qquad (2.1)$$

$$\text{s.t. } V(x,T) = D(X)$$

where $\dot{V}(x,t)$ is the Bellman value function (unknown scalar) or the cost incurred from starting in state x at time t and controlling the system optimally until time $T$. $C$ is the scalar cost rate function, $D$ is the final utility state function, $x(t)$ is the system state vector, $x(0)$ is an assumed given state. Being a minimization problem, the solution yielded from this equation is the value function or the minimum cost for a given dynamic system. The HJB equation is the classic method to solve optimal control problems, and to define a Reinforcement learning problem through Markov decision processes.

**Markov decision processes and Reinforcement Learning**

Markov decision processes (MDPs) are defined as time stochastic control processes based on the assumption of the Markov property, which states that given a present state, the future is independent of the past. This assumption allows us to affirm that the present state gives us the same information about the future as all the past states plus the present one. Mathematically, a state yields the Markov property if, and only if the following equation is satisfied:

$$P\left[S_t + 1 \mid S_t\right] = P\left[S_t + 1 \mid S_1, \ldots, S_t\right] \tag{2.2}$$

For this mathematical equation Markov processes are considered to be memory-less, and they basically randomly move from state to state. Markov processes can be defined with a tuple on a state space $S$ where states change via a transition function $P$:

$$P_{ss'} = P\left[S_{t+1} = s' \mid S_t = s\right] \tag{2.3}$$

where $S_t$ is a Markov state at time $t$, and $S_{t+1}$ is the next state. This transition function describes a probability distribution, where the distribution is the entirety of the possible states that the agent can transition to. Finally, there is a reward that is received for each state update, which is mathematically defined:

$$\begin{aligned} R_s &= E\left[R_{t+1} \mid S_t = S\right], \\ G_t &= R_{t+1} + \gamma R_{t+2} + \gamma^2 R_{t+3} + \cdots + \gamma^{k-1} R_{t+k} \end{aligned} \tag{2.4}$$

where $\gamma$ is the discount factor comprised between 0 and 1, $G_t$ is the total discounted reward and $R$ is the reward function. Therefore, a Markov reward process (MRP) can be defined with the tuple ($S$, $P$, $R$, $\gamma$). An example of an MDP with three states (green circles) and two actions (orange circles), with two rewards (orange arrows), and the probabilities of each action $a$ in state $s$ to lead to state $s'$ (in the black arrows), is visualized in Figure 2.1.

The agent can move from one state to another through an action receiving a reward, knowing that there is a certain probability of getting certain rewards for each action. The final goal for the agent is to learn to choose the process that accumulates the most rewards in a given episode given the parameters of the environment. This goal, in essence, is basically also the goal of a reinforcement learning process, where the strategy for choosing actions and maximizing the reward is defined in the so-called policy. A policy $\pi(s)$ is defined as the probability distribution of actions given a state $s$, mathematically defined as:

$$\pi\left(A_t = a \mid S_t = s\right), \forall A_t \in \mathcal{A}(s), S_t \in \mathcal{S} \tag{2.5}$$

Policies are grouped into two classes:

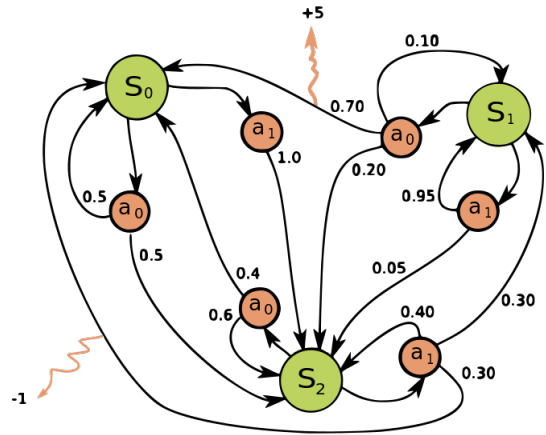

Figure 2.1: Markov decision process example. Source: Wikipedia (2022, September 29) [13]

- Stochastic: A policy that yields a probability distribution over a set of actions, such that there is a probability that the action taken will not be the action that occurs, usually associated with partially observable Markov processes (POMP).

- Deterministic: A policy that precisely maps a given state to an action (the policies used in the Gradient learning algorithms used in this research are all represented by a Neural Network (NN), which is deterministic once they have been trained).

A fundamental component of the Reinforcement Learning paradigm is learning by trial and error, a method well-studied in animal behavior. In other words, the reinforcement learning paradigm is based on the basic reward and punishment mechanisms that "reinforce" an optimal behavior. The variables seen in the MDP can be mapped into the RL coding variables as follows:

- Action ($A_t$): the action taken by the agent at time-step $t$ that will update the environment and generates a reward.

- Reward ($R_t$): the positive or negative reward given to the agent for the chosen action. Its value indicates the quality of action w.r.t. accomplishing the goal.

- Observation ($obs$): the State $S_t$ of the environment seen by the agent after an action has been performed.

- Done: a Boolean value that indicates whether the episode is finished. Once the episode is done, the environment needs to be reset.

- Info: a dictionary with miscellaneous information for debugging.

These variables define the learning paradigm scheme used in RL environments as can be seen in Figure 2.2. The graph shows the cycle of a time-step $t$ where the environment observation state $S_t$ and its reward $R_t$ trigger the agent's action $A_t$, which updates the environment to a new state $S_{t+1}$ with a new reward $R_{t+1}$. The cycle continues until a stopping condition sets the $Done$ variable to $True$; then, the environment resets its state $S_t$ to the initial state.

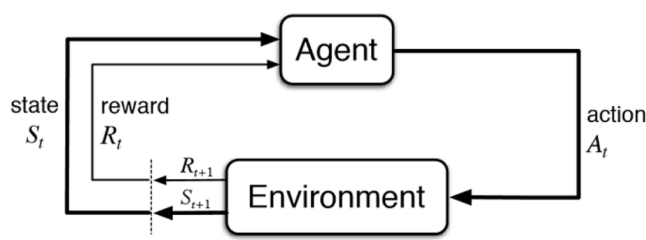

Figure 2.2: Reinforcement learning paradigm.

## 2.3 The Reinforcement Learning framework

This section describes the tools used to develop this research from an RL project point of view. The code backing up this research was indeed developed on a framework suitable for RL custom environments and training. The framework used consists of:

- The RL Environment, built with the OpenAI Gym library.

- The RL Agent model, built using Stable Baselines3 models based on different learning algorithms.

- The model training and tuning platform: Weights  Biases.

Each framework instance is built to work in Python3 and it integrates with Jupyter notebooks, where most of the operations for this project are done.

**OpenAI Gym**

[14] The OpenAI Gym is a Python library that provides several environments in which users can begin utilizing RL algorithms and they can easily start training their RL models. OpenAI Gym is famous for implementing several vintage games in the Python framework where a trained RL agent can play. The library is also very useful and famous for creating a standard structure to build a **custom RL environment**, that can work for any purpose. The RL custom environment built for this thesis project, the VPP environment ($VPPenv$) is based on the OpenAI Gym structure. The core component Gym offers is indeed the environment. A complete description of the VPP environment is described in the next chapter. However, for the environment definition and the model algorithm implementation chapters, it is important to note that the Gym action space class ($gym.spaces$) proposes four main different types of spaces:

- Box: A N-dimensional box that contains every point in the action space.

- Discrete: A list of possible actions, where each time-step only one of the actions can be used.

- Multi-Discrete: A list of possible actions, where each time-step only one action of each discrete set can be used.

- Multi-Binary: A list of possible actions, where each time-step any of the actions can be used in any combination.

The Observation spaces implement also:

- Dict: A dictionary of all the observation variables that can be any of the types listed above.

**Stable Baselines3**

Stable Baselines3 (SB3) [15] is a set of reliable implementations of reinforcement learning algorithms in PyTorch [16]. SB3 is implemented in **PyTorch**, a famous optimized tensor library for machine learning. SB3 provides a reliable RL algorithm library with models ready to use, focusing on model-free, single-agent RL algorithms and offline learning. The main tools used by SB3 are:

- the Vectorized and Monitor environment wrappers, which allow, respectively, parallelizing several models' training and monitoring their training parameters.

- The Environment check function, which tests if the environment is compliant with the OpenAI gym environment rules and therefore the SB3 algorithms.

- The model evaluation function, which allows parallel tests on the trained model to evaluate the average cumulative reward and standard deviation.

- The A2C and PPO model algorithms.

- The MaskablePPO, TRPO, and RecurrentPPO model algorithms, taken from the Stable-Baselines3-contrib branch [17].

The RL algorithms are chosen with the features: single process, Multi-Discrete action space. The SB3 model will be coupled with the Weights and Biases (Wandb) interface to track the development of the project with the different training sessions and to visualize progress.

**Weights & Biases**

Weights & Biases (W&B) [18] is a machine learning platform geared for ML project development. It is designed to support and automate key steps in the Machine Learning Operations (MLOps) life cycle, such as experiment tracking, data-set versioning, and model management. The platform connects directly from the Jupyter notebook's API through your credentials, linking all the data generated during that run in your project dashboard. The main tools used by WB are:

- the W&B Callback function, to get callbacks while training and storing data.

- The W&B documents generation and the GUI Visualization tables.

- The W&B Hyperparameter Sweeps, for the Hyper-parameters tuning.

**Google Colaboratory (Colab)**

The main coding platform used is Google Colaboratory (Colab) [19], which is a famous data analysis and machine learning tool that allows to combine executable Python code and rich text along with interactive charts, images, and more into a single document stored in Google Drive. It uses Jupyter notebooks run on remote Ubuntu 18.04 servers also available with GPUs and TPUs for training AI models. The computations and results proposed in this research are entirely deducted from the Colab notebooks of a Drive repository.

## 2.4 Reinforcement learning Algorithms theory

This section presents and describes the theoretical background of the algorithms used to train the RL model of this research. The following algorithms are selected from the Stable Baselines3 library, because of their acquired notoriety in the literature, and their great support from the community. Furthermore, the algorithms proposed were chosen because of their feasibility for the custom environment used. Specifically, the action space and the observation space shapes used, which require the support for the Multi-Discrete action space and the "Dict" observation space. While the latter is widely adopted in most of the SB3 algorithms, the former has currently support only for the A2C, PPO algorithms and alike. Both of them are part of the policy-based learning algorithm family.

### 2.4.1 Policy-Gradient learning algorithms overview

In RL theory, [12] policy-based gradient learning methods are focused on optimizing the policy function, i.e. the action selection strategy. The opposite approach used by value-function learning methods is instead trying to learn a value function, that is, a function yielding information on the expected rewards in a given state. Basically, gradient methods point toward the steepest point of the function to differentiate. If applied to an error function, and used in the form of gradient descent, it will adjust the action selections to minimize the error function's value (locally or globally). As such, gradient methods are generally reaching a feasible solution. The main features of policy gradient methods are:

- Gradient methods are particularly adept at learning stochastic processes whereas value-based functions cannot.

- Gradient methods are guided toward a solution by a gradient, which leads them to converge faster on solutions than value-based methods.

- Gradient methods do not have an exploration/exploitation trade-off, so they do not need to explore an environment by taking the same action many times.

- Gradient methods are significantly more effective in high-dimensional spaces because they are not computationally expensive. They just perform an action and adjust the gradient.

In contrast, value-based methods can yield a considerably larger and more non-intuitive range of values between actions of minimal difference, which does not guarantee convergence. Value-based methods calculate a value for each possible action, which makes practically impossible the convergence on a space with a considerably high number of actions.

### 2.4.2 Policy Gradients mathematical explanation

The objective of a Reinforcement Learning agent is to maximize the "expected" reward when following a policy $\pi$. [20] Defining the total reward for a given trajectory $\tau$, as $r(\tau)$, it can be defined as follow:

$$\arg\max J(\theta) = E_\pi[r(\tau)] \qquad (2.6)$$

Where $\theta$ is the set of parameters that parameterize the policy (e.g. neural network's weights and biases or polynomial coefficients). For a given finite MDP, it always exists at least one optimal policy that gives a maximum reward, that is also stationary and deterministic. To find the optimal policy is only necessary to get the parameters $\theta^*$ (optimal set of parameters $\theta$). The standard approach to solve a minimization/maximization problem is to use gradient descent (or ascent). The gradient ascent method tries to get closer to desired values through steps following the updating rule:

$$\theta_{t+1} = \theta_t + \alpha \nabla J(\theta_t) \qquad (2.7)$$

Therefore, the objective of the reward maximization problem is defined as follows:

$$\theta^* = \arg\max_{\theta} E_{\pi\theta} \left[ \sum_t R(s_t, a_t) \right] \qquad (2.8)$$

This function tries to pick the optimal parameters that maximize the reward yielded for actions taken within a given state, which in this case, they are the weights for the network that maximizes the score. The derivative of the expected total reward is therefore defined as follows:

$$\nabla E\pi[\mathrm{r}(\tau)] = E\pi[\mathrm{r}(\tau) \nabla \log \pi(\tau)] \qquad (2.9)$$

This last equation is obtained through the policy gradient theorem, which states that the derivative of the expected reward is the expectation of the product of the reward and gradient of the log of the policy $\pi_\theta$. Then, expanding the definition of $\pi_\theta$:

$$\pi(\tau) = P(s_0) \prod_{t=1}^{T} \pi_\theta(a_t \mid s_t) \, p(s_{t+1}, r_{t+1} \mid s_t, a_t) \qquad (2.10)$$

where $P$ is the ergodic distribution of starting in some state $s_0$. Then the product rule of probability applies because each new action probability is independent from the previous one for the definition of the MDP itself. At each time-step, an action is taken by the policy $\pi_0$ and the environment dynamics $p$ decide which new state to transition into. In the end, there are $T$ terms, equal to the number of time-steps representing the length of the trajectory. By taking the logarithm of the expected reward, the equation can be rewritten as a sum as follows:

$$\log \pi_\theta(\tau) = \log \mathcal{P}(s_0) + \sum_{t=1}^{T} \log \pi_\theta(a_t \mid s_t) + \sum_{t=1}^{T} \log p(s_{t+1}, r_{t+1} \mid s_t, a_t)$$

$$\nabla \log \pi_\theta(\tau) = \sum_{t=1}^{T} \nabla \log \pi_\theta(a_t \mid s_t) \tag{2.11}$$

$$\nabla E_{\pi_\theta}[r(\tau)] = E_{\pi_\theta}\left[r(\tau)\left(\sum_{t=1}^{T} \nabla \log \pi_\theta(a_t \mid s_t)\right)\right]$$

What this result means is that, surprisingly, the knowledge about the ergodic distribution of states $P$ and the environment dynamics $p$ is not needed for the expected reward. This is a very convenient demonstration since it is hard to model both of these variables. For this reason, all the algorithms that use this result are called model-free algorithms, because they do not need to "model" the environment. The only thing that they are modeling is the rewards.

### 2.4.3 Advantage Actor-Critic models (A2C)

One of the most advanced models available in the literature are the Actor-Critic Models [21], which are characterized by the fact that they update their parameters for all of the models simultaneously. These models learn from action to action, rather than for every episode like many other Reinforcement Learning algorithms. This step-wise rather than episodic update is what characterizes the Actor-Critic models, on which A2C and PPO are based. The Actor-Critic models are characterized by the advantage factor, defined as follows:

$$A(s_t, a_t) = r_{t+1} + \gamma V_v(s_{t+1}) - V_v(s_t) \tag{2.12}$$

where $V_v(s_t)$ is the average value of states $s_t$. This value function estimate behaves like a "critic" (it evaluates good against bad values) to the "actor" (the agent's policy). The convenient thing for this factor is that only one neural network for the $V$ function (parameterized by $v$) is needed, compared to other methods. The advantage factor inserted in the "vanilla" policy gradient equation become:

$$\begin{aligned}
\nabla_\theta J(\theta) &\sim \sum_{t=0}^{T-1} \nabla_\theta \log \pi_\theta(a_t \mid s_t)(r_{t+1} + \gamma V_v(s_{t+1}) - V_v(s_t)) \\
&= \sum_{t=0}^{T-1} \nabla_\theta \log \pi_\theta(a_t \mid s_t) A(s_t, a_t)
\end{aligned} \tag{2.13}$$

This equation is mathematically equal to the same used by the A2C (Advantage Actor-Critic) algorithm.

### 2.4.4 Proximal policy optimization (PPO)

PPO is the most famous algorithm in the RL literature as it overcomes the policy gradient tendencies of getting stuck in local minima/maxima. It can do so by imposing a penalty on the objective function and then using this newly reformed gradient descent [22]. The PPO algorithm also introduces the policy ratio, i.e. the probability ratio between the new policy and old policy defined as $r(\theta)$:

$$r(\theta) = \frac{\pi_\theta(a \mid s)}{\pi_{\theta old}(a \mid s)} \quad (2.14)$$

[23] PPO imposes the policy ratio to stay within a narrow interval between $1 - \epsilon$ and $1 + \epsilon$. The variable $\epsilon$ is an hyper-parameter called "clipping parameter" or epsilon, which is set by default in PPO vanilla implementations to 0.2. With the policy ratio implementation, the PPO equation is:

$$\max_\theta \widehat{E}_t \left[ \frac{\pi_\theta(a_t \mid s_t)}{\pi_{\theta_{\text{out}}}(a_t \mid s_t)} \hat{A}_t \right] - \beta \widehat{E}_t \left[ KL \left[ \pi_{\theta_{ad\theta}}(\cdot \mid s_t), \pi_\theta(\cdot \mid s_t) \right] \right] \quad (2.15)$$

where $\beta$ is the tuning parameter, $KL$ is the Kullback–Leibler divergence, $A_t$ is the advantage function seen in the A2C algorithm. The adaptive penalty works using the $KL$ divergence between the old and the new policy, which changes every time-step during an episode. The value from the $KL$ divergence serves to tune $\beta$ and the search region of the algorithm. This search region is the same "trust region" described for the TRPO algorithm. The benefit of adding the penalty is that it ensures that the area in which the algorithm looks for the parameters to define the policy is significantly smaller. Furthermore, it adjusts the trajectory update on a step-wise level rather than an episodic one. This way, bad actions are directly penalized instead of being averaged out with the whole action series.

### 2.4.5 Maskable proximal policy optimization (MaskablePPO)

[24] Invalid action masking is a technique used in Reinforcement Learning to avoid repetitive invalid action generation in large Multi-discrete action spaces, and ultimately to speed up the agent learning. The alternative commonly used for invalid action masking is the invalid action penalty, which consists in attributing

a negative reward when an invalid action is chosen. However, this method is not quick for policy optimization. The policy gradient algorithms studied in this research, employ a neural network to represent the policy. The neural networks' output consists of unnormalized scores for each possible action (logits) that are then converted into action probability distribution using a softmax operation. Thus, the invalid action mask is applied by replacing the logits corresponding to the invalid actions with negative infinity, before passing the logits to the softmax function.

An example of action masking for a Multi-discrete (Multi-binary in this case) action space represented as $[2, 2]$, can be explained as follows, by using the same notation of MDP and gradient policy variables. The action set for such space can be defined as $A = \{a_0, a_1, a_2, a_3\}$, and the policy $\pi_\theta$ parameterized by $\theta = [l_0, l_1, l_2, l_3] = [1.0, 1.0, 1.0, 1.0]$. action $a_0$ corresponds to choosing 0 for the first discrete value, action $a_1$ chooses 1 for the first discrete value, action $a_2$ chooses 0 for the second discrete value and action $a_3$ chooses 1 for the second discrete value. By default, the probability of choosing any of these actions is equal for all of them. Now, supposing $a_2$ is invalid for state $s_0$, the valid actions are $a_0, a_1, a_3$. The action mask is applied by replacing the logits of the actions to be masked by a large negative number $M$ (e.g. $M = -1 \times 10^9$). The re-normalized probability distribution $\pi'_\theta(\cdot \mid s_0)$ is evaluated as follows:

$$
\begin{aligned}
\pi'_\theta(\cdot \mid s_0) &= \text{softmax}\left(\text{mask}\left([l_0, l_1, l_2, l_3]\right)\right) \\
&= \text{softmax}\left([l_0, l_1, M, l_3]\right) \\
&= [\pi'_\theta(a_0 \mid s_0), \pi'_\theta(a_1 \mid s_0), \epsilon, \pi'_\theta(a_3 \mid s_0)] \\
&= [0.33, 0.33, 0.0000, 0.33]
\end{aligned}
\quad (2.16)
$$

where $\epsilon$ is the resulting probability of the masked invalid action, which should be a small number. If $M$ is chosen to be sufficiently negative, the probability of choosing the masked invalid action $a_2$ will be virtually zero. the invalid action masking at this point re-normalized the probability distribution of the actions. Thus, the invalid action policy gradient is evaluated such that as a result, the logit of the invalid action is equal to zero:

$$
\begin{aligned}
g_{\text{invalid action policy}} &= E_\tau \left[\nabla_\theta \sum_{t=0}^{T-1} \log \pi'_\theta(a_t \mid s_t) G_t\right] \\
&= \nabla_\theta \log \pi'_\theta(a_0 \mid s_0) G_0 \\
&= [0.67, -0.33, 0.0000, -0.33]
\end{aligned}
\quad (2.17)
$$

## 2.4.6 Trust region policy optimization (TRPO)

[25] Many Gradient descent/ascent method uses line search as an improving tool to find the global maxima for the optimization task. The line search method firstly finds the step size $\alpha$, and secondly determines the direction of the gradient. The trust region method instead, first decides the step size $\alpha$, then it marks a circular region by considering $\alpha$ as the radius of the circle, that is, the trust region. Then, the research for the local minimum or local maximum of the problem is bounded to that region. The optimal point found in that region determines the direction.

[22] TRPO introduces trust region strategies to RL, substituting the line search method. TRPO uses the $KL$ divergence constraints to create the trust-region for the optimization process. TRPO ensures that every policy update is not far from the previous one by keeping the new policy within the trust region of the old policy. The objective function of the TRPO algorithm is:

$$E_{\tau \sim \pi_{\theta_{old}}} \left[ \frac{\pi_\theta(a \mid s)}{\pi_{\theta_{old}}(a \mid s)} A_{\theta_{old}}(s, a) \right] \quad (2.18)$$
$$\text{subject to } E_{\tau \sim \pi_{\theta_{old}}} \left[ \overline{D_{KL}} \left( \pi_{\theta_{old}}(. \mid s), \pi_\theta(. \mid s) \right) \right] \leq \delta$$

The equation can be interpreted as a maximization of the objective function subject to the $KL$ divergence constraint term on the step size of the policy update. This mathematical constraint makes the TRPO algorithm relatively complicated as the KL constraint adds additional overhead in the form of hard constraints to the optimization process [26]. The PPO algorithm is a much simpler approach since it is its first-order implementation.

## 2.4.7 Recurrent proximal policy optimization (RecurrentPPO)

The RecurrentPPO is an implementation of the Proximal Policy Optimization (PPO) algorithm with support for recurrent policies. The Recurrent PPO is also known as LSTM PPO (Long-Short-Term-Memory PPO) since it introduces the hidden states support to enable memory in the observation changes through different time-steps. Other than adding support for recurrent policies, the behavior is the same as in SB3's core PPO algorithm. The recurrent policies are simply policies that use Long-Short-Term Memory Networks. [27] LSTM networks are a special kind of RNN (Recurrent Neural Network). They are characterized by the parts composing the single LSTM cell:

- The cell state: the aggregated or global memory of the LSTM network for all time-steps.

- The hidden state: an encoding of the most recent time-step. It can be processed to obtain more meaningful data.

- The gates: divided in input, output, and "remember"; they perform feature-extraction to encode the data that is meaningful to the LSTM and determining how memory-important data are. The feature-extracted matrix is then scaled by its memory-importance before getting added to the cell state, or in other words the global memory of the LSTM network.

- The hidden size (num_units): this size is analogous to the number of "neurons" (perceptrons) in the Dense Neural Networks (DNNs) context, in a given layer of the network. A visualization of the hidden units of connected LSTMs cells is shown in Figure 2.3.

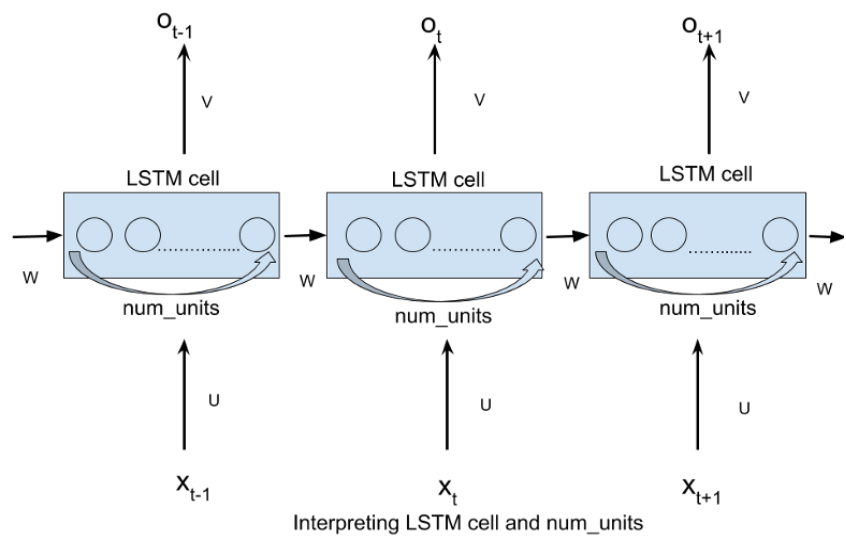

Figure 2.3: Graphic illustrating hidden units within LSTM cells. Source: Medium.com [27]

To implement a RecurrentPPO algorithm model, the details for integrating LSTM support are [28]:

- LSTM Neural Network layers initialization: The weights are randomly initialized with a standard deviation equal to 1, and the biases set to 0.

- LSTM states initialization: all the LSTM states (hidden and cell states) are initialized to zeros in the ($initial\_lstm\_state$).

- LSTM states reset: after the end of every episode, during rollouts or training, the "Done" flag is checked to reset The LSTM states with zeros.

- LSTM states reconstruction while training: the algorithm model saves a copy of the LSTM states ($lstm\_state$) for every step. During the episode, the agent sequentially reconstructs the LSTM states based on the ($initial\_lstm\_state$) fed at the beginning. This process ensures that the probability distributions are successfully reconstructed in the rollouts.

## Chapter 3

# Pre-processing and analysis

This chapter focuses on the analysis made for this research. Specifically, the analysis of the beginning state of work and the steps to complete to have a working environment. It continues with the description of the data-sets used, the pre-processing applied, and the subdivision into training, testing, and validating data-sets.

## 3.1 The starting point: Elvis

Each year of a VPP (virtual power plant) simulation is based on one year of an Elvis simulation. Elvis [29] is a planning and management tool for electric vehicle charging infrastructure developed at DAI-Labor [30]. It enables the modeling of the stochastic nature of wind and solar power generation and demand fluctuation of customer agents, implementing demand-side management programs for customer agents. It also works as an infrastructure planner simulating charging behavior, arrival distributions, vehicle specs, and electric rates. Elvis simulation's features will be the focus of the VPP starting point.

Elvis' most useful feature for this research is the generation of charging events: Elvis simulation generates a list of EV charging events within the time-range selected, specifying arrival time, departure time, state of charge (SoC) at arrival, and the Vehicle type. From the charging events list, it is possible to evaluate the EVs' load over time created through **uncontrolled charging**. By uncontrolled charging, it is meant that the EVs connect and start charging at maximum power until they reach the maximum capacity. Each vehicle in the simulation has its own ID represented by an integer number, starting from 1 up to 2147483647 (the maximum int32 value).

The Elvis simulation is already a great starting point for the design of a VPP simulator. Elvis generates pseudo-random arrivals and departures of vehicles with a certain SoC according to the parameters set in the Elvis configuration file. An example of the data included in the Elvis configuration file is visualized in Figure 3.1. It is important to note that the average parking time limit for each EV in the simulation is 24 hours.

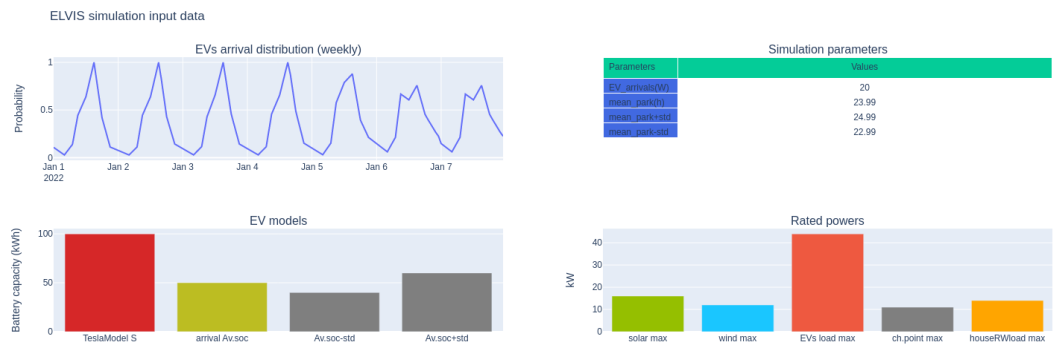

Figure 3.1: Example visualization of the Elvis configuration file.

The EVs arrival time is a random variable with a weekly probability distribution defined in the Elvis configuration, like the example probability function displayed in the top-left corner of Figure 3.1. The time resolution of the arrival probability distribution is the same as the Elvis simulation defined in the configuration. The parking time and the SoC at arrival of each EV are instead random variables with a gaussian probability distribution with mean and standard deviation defined in the Elvis configuration as shown in the top-right table of Figure 3.1. An example of an arrival SoC distribution of an Elvis simulation is shown in Figure 3.2.

However, the Virtual Power Plant's goal is not solely to charge the vehicles as fast as possible, but to provide support to the grid by charging or discharging vehicles when needed. Therefore, the arrival and departure times of EVs will be kept for the VPP simulation, but the EVs load profile (generated by uncontrolled charging) will be used only for comparison between controlled and uncontrolled charging loads. Thus, for each episode of the VPP simulation (1 year long) a new Elvis simulation is launched to update the charging events list. This charging list is accessed throughout the whole simulation to define how many vehicles are present for each time-step. This state is expressed by the EVs IDs array. The description of the VPP simulation states is presented in the next section.

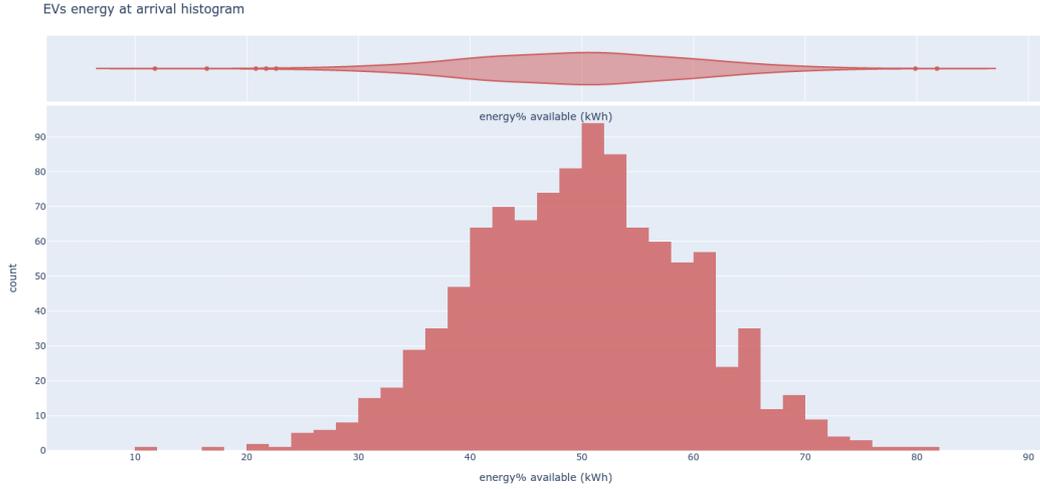

Figure 3.2: Example distribution of an Elvis simulation EVs available energy at arrival.

## 3.2 The RL framework

The VPP simulation was set up following the Reinforcement Learning framework explained in the previous chapter. The RL variables are all addressed in the VPP environment cycle as shown in Figure 3.3, following its own update rules. The represented simulation scheme follows the RL paradigm: for each time-step $t$, the environment observation state $S_t$ and its reward $R_t$ trigger the agent's action $A_t$, which updates the environment to a new state $S_{t+1}$ with a new reward $R_{t+1}$.

**The Environment state spaces**

The first scheme of the VPP framework was based on the fact that the agent needed as much significant state data as possible to predict the best action. The observation space included five different state variables, two of which were arrays with the size of the charging station number defined in the Elvis simulation. The observation space is expressed in the environment code as a Python dictionary

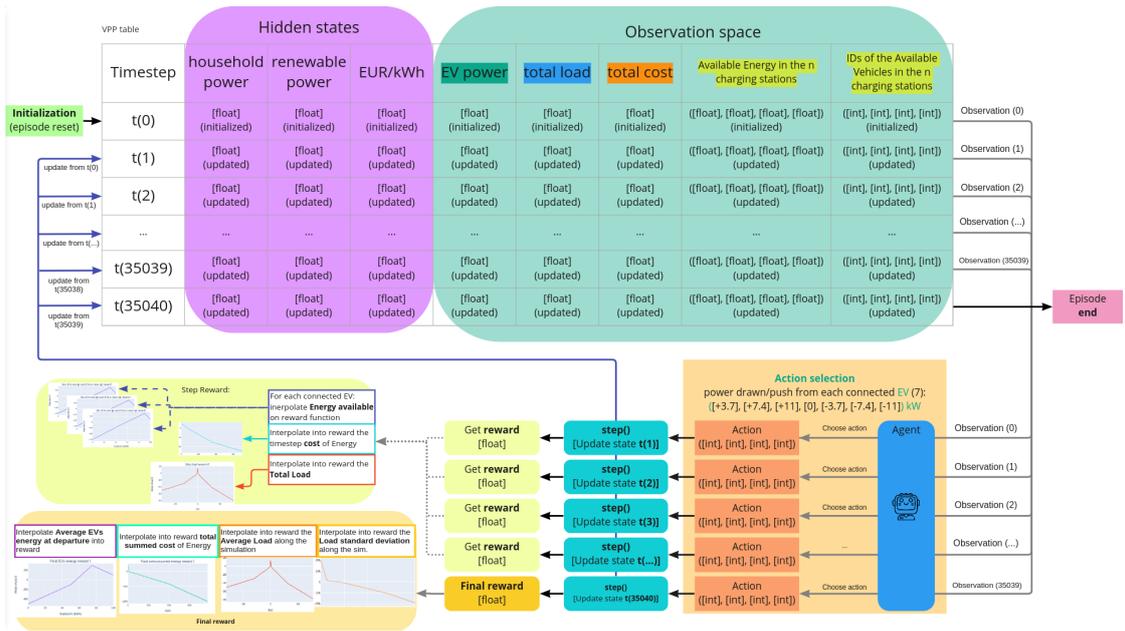

Figure 3.3: The first idealization of the VPP RL framework structure.

defining the state variables that can be described as follows:

- EV power: a Box type (continuous) scalar indicating the total power drawn/pushed from the EVs at time-step $t$.

    - It is evaluated by summing the actual power drawn/pushed from/to the connected EVs.

    - It ranges from the maximum charging station power defined in the Elvis config file and its negative value.

    - It depends on the actions chosen by the agent, and by the Available energies of the EV connected to the charging stations' state.

- Total load: a Box type (continuous) scalar indicating the resulting power from the supply/demand sum at time-step $t$.

- It is evaluated every time-step $t$ by the formula:

$$total\_load_t = households\_load_t + renewable\_load_t + EVs\_load_t \tag{3.1}$$

- It ranges from the maximum households' consumption plus the maximum charging station power defined in the Elvis configuration file and its negative value. Renewable power has negative values since it is supplied power while EV power can be positive or negative depending on the charging/discharging action.

- Total cost: a Box type (continuous) scalar indicating the overall cost of the energy from the Grid consumed at time-step $t$.

  - It is evaluated every time-step $t$ by the formula (converting to kWh, with time-step $\delta t$ being 15 minutes or $\frac{hour}{4}$):

$$total\_cost_t = total\_load_t \cdot \delta t \cdot energy\_price_t \tag{3.2}$$

  - It ranges from the max Total load value divided by 4 times the highest value in the energy prices series of the input data-set, and its negative value.

- Available energies: an array of $n$ Box type (continuous) scalars of the available energies at the charging stations, $n$ being the number of charging stations in the simulation.

  - Each array value ranges from 0 (no EVs connected) to 100 kWh (max EVs simulation capacity). Minimum feasible charge if an EV is connected is 0.1 kWh.
  - It depends on the Elvis simulation charging events list and on the actions taken in the past (charging/discharging connected EVs battery).

- Available EV IDs: an array of $n$ integer values indicating the available EV IDs at the charging stations, $n$ being the number of charging stations in the simulation.

– Each array value ranges from 0 (no EV connected) to 2147483647 (max int32 value).

– It is entirely dependent on the Elvis simulation charging events list.

There are other hidden state variables in the VPP simulation that are not fed to the agent in the observation space: household power, renewable power, and energy market prices. These time-series come from the input data-set (different series for training, testing, and validating data-set) and are not directly accessible by the agent, but they are indirectly extractable from the observation space variables. The agent shall learn the patterns of these hidden variables by correlation with the *total_load* and *total_cost* spaces.

**The action space**

The action space is the set of actions the agent can potentially choose each time-step. It corresponds to the decision of the agent towards the connected vehicle power direction. The first idealization of the action space consisted of a Multi-Discrete space with 7 actions per charging station. Each action per charging station, numbered from 0 to 6, has its own power transfer direction and intensity (converting to kWh, with time-step $\delta t$ being 15 minutes or $\frac{hour}{4}$):

- Action 0: Idle. No power drawn/pushed from the EV;
- Action 1: Charge [+3.7 kW for $\delta t$] the connected EV by 0.925 kWh;
- Action 2: Discharge [-3.7 kW for $\delta t$] the connected EV by 0.925 kWh;
- Action 3: Charge [+7.4 kW for $\delta t$] the connected EV by 1.85 kWh;
- Action 4: Discharge [-7.4 kW for $\delta t$] the connected EV by 1.85 kWh;
- Action 5: Charge [+11 kW for $\delta t$] the connected EV by 2.75 kWh;
- Action 6: Discharge [-11 kW for $\delta t$] the connected EV by 2.75 kWh;

The EV load state variable is evaluated by summing the power chosen by the agent for each station. The Agent must pick an action for each charging station. The actions' effects are subject to some restrictions.

- Actions different from 0 (Idle) taken for a charging station at time-step $t$ without an EV plugged-in are useless and do not change any result.

- Charging an EV with a resultant energy greater than 99.9 kWh does not produce any effect (the excess energy over the 99.9 kWh is reconverted in power and subtracted from the EV power).

- Discharging an EV with a resultant energy smaller than 0.1 kWh does not produce any effect (the excess energy surpassing the minimum threshold of 0.1 kWh is reconverted in power and subtracted from the EV power).

- EVs connected with a SoC below 10% (10 kWh in this environment) cannot be discharged or kept idle. They will be charged instead with Action 1 if Action 3, or Action 5 were not selected for that EV.

- EVs connected with a SoC below 20% (20 kWh in this environment) cannot be discharged. They will be kept Idle instead if Action 1, Action 3 or Action 5 were not selected for that EV.

These rules are the same that will asses whether the action taken was valid or invalid. The actions' validity concept is used in the Action Masking learning approach explained in the next chapters. It is possible to identify each possible action in a numbered table as shown in Figure 3.4.

| Action codes table | | Charging station $n$ | | | |
|---|---|---|---|---|---|
| Action meaning | Action Value | 0 | 1 | 2 | 3 |
| Idle [+0 kW] | 0 | 0 | 1 | 2 | 3 |
| Charge [+3.7 kW] | 1 | 4 | 5 | 6 | 7 |
| Discharge [-3.7 kW] | 2 | 8 | 9 | 10 | 11 |
| Charge [+7.4 kW] | 3 | 12 | 13 | 14 | 15 |
| Discharge [-7.4 kW] | 4 | 16 | 17 | 18 | 19 |
| Charge [+11 kW] | 5 | 20 | 21 | 22 | 23 |
| Discharge [-11 kW] | 6 | 24 | 25 | 26 | 27 |

Figure 3.4: The first idealization of the VPP action space.

In this visualization, actions can be numbered from 0 up to 27, evaluated through the equation:

$$action\_number = ((charging\_station\_n + 1) \cdot (action\_value + 1)) - 1 \quad (3.3)$$

Where *charging_station_n* is the number of the charging station where the action is applied (ranging from 0 to 3), and the *action_value* is the code of the action to be applied (ranging from 0 to 6). The last action number is therefore evaluated with the formula:

$$action\_number_{max} = charging\_station\_N \cdot action\_values\_N - 1 \qquad (3.4)$$

Where *charging_station_N* is the number of charging stations (4, in this case) and *action_values* is the number of different action values (7, in this case). This action labeling for each station serves only the purpose of action masking and it is irrelevant to the general action space definition. In any case, the action vector generated by the model is an array of 4 numbers comprised between 0 and 6. Since for each time-step the agent must pick an action for each charging station (4 in the studied environment), the action set is composed of 2401 action combinations ($7^4$).

**Reward evaluation**

The VPP simulation evaluates a reward for the observation space for each time-step. An additional final reward at the last step of the simulation is added to the reward evaluated for the last step. A focus from the RL framework on the reward functions is available in Figure 3.5.

In the first idealization of the VPP RL framework, the reward for each time-step is evaluated considering 2 observation state variables:

- The reward attributed to the total load value at time-step $t$: The total load value is mapped into an interpolated reward function peaked at 0 kW, positive between -1.5 kW and 1 kW and negative otherwise.

- The reward attributed to the available energies at the charging stations at time-step $t$: For each connected EV, the available energy percentage value is mapped into an interpolated reward function positive after 55% peaked at 90%. If for charging station $n$ there is no EV connected, no reward is evaluated.

- The reward attributed to the available energy at the departure of each EV. This reward is evaluated exclusively when an EV leaves the charging station, and its available energy percentage value is mapped into an interpolated

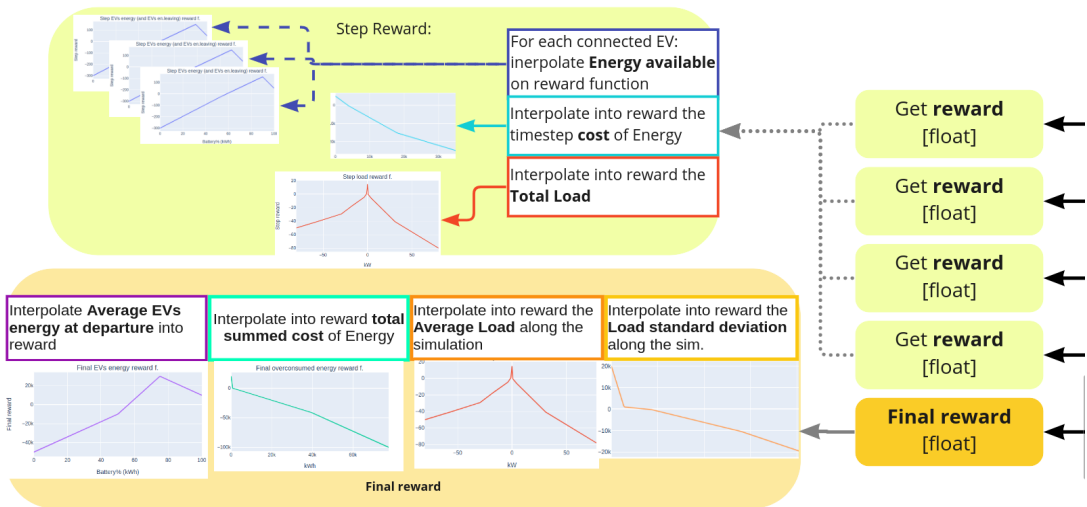

Figure 3.5: Focus on the rewards evaluation from the first implemented RL framework

reward function with the same shape as the previously described one.

The final reward is instead evaluated at the end of the simulation and it considers different parameters extrapolated once the simulation is complete:

- The reward attributed to the Average EVs available energy at departure: the average energy percentage value is mapped into an interpolated reward function positive after 55% and peaked at the maximum average EVs available energy at departure, a parameter described in the next section.

- The reward attributed to the total sum of the load values for all simulation time-steps: The summed total load value is mapped into an interpolated reward function peaked at 0 kW, positive between plus/minus half of the summed total load value of the Elvis simulation, and negative otherwise.

- The reward attributed to the average load for all simulation time-steps: The average load value is mapped into an interpolated reward function peaked at 0 kW, positive between -1.5 kW and 1 kW, and negative otherwise.

- The reward attributed to the standard deviation of the load for all simulation time-steps: The standard deviation load value is mapped into an interpolated reward function peaked at 0 kW, positive between 0 and half of the standard deviation load of the Elvis simulation and negative onward.

## 3.3 The data

For a truthful VPP environment development, the necessary data to be gathered are:

- the photovoltaic (PV) solar power generation over time profile;

- the wind turbines (WT) power generation over time profile;

- the Household power consumption profile over time;

- the energy market prices over time.

The time-series of data must have a resolution of 15 minutes to allow a detailed simulation.

### 3.3.1 The data-sets source

All the data must have the same origin place of sampling and time-span of one year, possibly during the same year-frame. A Data set including data-series from the same place but with one or two years of difference are still considered valid. To satisfy these requirements, the following open-source data portals or data generation tools were taken into account:

- Renewables Ninja [31] for the wind and solar power profiles generated through a simulation of hourly power output from wind and solar PV farms in the indicated place and using a specific technology. The data-set used for the simulation is the MERRA-2 global from the years 2018, 2019, and 2020.

- Pecan street Dataport [32] for the energy market prices. the chosen data series were the Day-ahead prices from the Entsoe [33] transparency platform. The data were exported from 2018, 2019, and 2020 energy market prices of Germany with a resolution of 1 hour or 15 minutes.

- Low-Voltage Load Forecasting[34] for realistic household load profiles over time. The chosen data-set is the BLEMdataGlimpse, a data-set containing time series of energy consumer/prosumer's electricity readings and timestamps from 2017 in Germany with a resolution of 3 minutes.[35]

Each raw dataset downloaded from its source was collocated in the "raw_datasets" folder of the training/testing/validating repository to be processed.

### Solar energy production infrastructure

The PV solar power generation infrastructure is simulated on Renewables Ninja according to the following technology and scenario settings:

- Coordinates of simulation: latitude: 51.0834196, longitude: 10.4234469 (Berlin).
- Dataset generator: Merra2.
- System loss: 0.1
- axes tracking enabled: 2. Tilt: 35 degrees; Azimuth: 180 degrees.
- Panels units: 40
- Single unit: 1.88m x 1m = 1.88 $m^2$. Total area: 60 $m^2$
- Single unit capacity: 400 W. Total capacity: 16 kW.

### Wind energy production infrastructure

The WT wind power generation infrastructure is simulated on Renewables Ninja according to the following technology and scenario settings:

- Coordinates of simulation: latitude: 51.0834196, longitude: 10.4234469 (Berlin).

- Dataset generator: Merra2.

- Turbine model: Automaxx WINDMILL

- Hub height: 80 meters.

- Turbine units: 8

- Blade diameter: 1.7 $m$

- Rated wind speed: 15 $m/s$

- Single unit capacity: 1500 W. Total capacity: 12 kW.

### 3.3.2 Data-sets pre-processing

The pre-processing of the raw data-sets consisted in loading the *csv* files in a Python Pandas data-frame and performing routine operations of data processing:

- Checking that no empty data-cells are present.

- In case of empty cells or unreadable data, forward-filling all invalid data-cells.

- Making sure the first and last time-steps coincide with the beginning and end of a year (e.g. 00:00 01-01-2018 – 00:00 01-01-2019).

- Re-sampling data to 15 minutes time-steps ($\delta t$).

- Checking that all the time-series have 35041 time-steps (365 days times 24 hours per day, times 4 quarters per hour, plus the first time step of the next year (00:00 01-01-2019/2020/2021).

- Setting the date-time series as the index of the Dataframe.

- Plotting time-series data to verify the feasibility of the profiles.

To ensure an equal number of time-steps for each data-set, since 2020 was a leap year, the testing data-set generated from the 2020 time-series was modified by deleting all the time-steps belonging to the 29th of February. Once the time-series are pre-processed and saved in the "scenario_datasets" folder as *csv* files, they are ready to be merged into the VPP input data-set.

### 3.3.3 VPP input dataset tables

Three VPP input data-sets are generated from the time-series:

- The training data-set, built from the 2019 series of Wind and Solar power and energy price series, with the households consumption data generated by the sum of four different household profiles from the BLEM datasets of 2017.

- The training data-set, built from the 2019 series of Wind and Solar power and energy price series, with the households consumption data generated by the sum of other four different household profiles from the BLEM datasets of 2017.

- The training data-set, built from the 2019 series of Wind and Solar power and energy price series, with the households consumption data generated by the sum of other four different household profiles from the BLEM datasets of 2017.

Each time-series from the pre-processed data (solar power, wind power, household load, and energy price) are added as columns of the VPP table (for training, testing, and validating). An example of what the content of a VPP input data-set should look like is shown in Figure 3.6. The generated topologies are used as input for the VPP environment class initialization.

**Data-sets noise alteration**

An important detail of the VPP simulation is that for each episode, white gaussian noise is applied to the renewable generated power and to the energy market prices time-series. The applied noise has zero mean and standard deviation equal to 10% of the greatest value of the time-series to be applied to. For each episode, the noise is applied to a copy of the initial data-set values so that the original data is preserved along the episodes. This alteration of the data for each episode is fundamental for the learning phase of the RL agent because it permits the extrapolation of the series patterns. The altered data-set is used by the VPP environment that will iterate through its time-series during the simulation. Once the VPP environment is initialized, the input data-set can be visualized as shown in Figure 4.7a, which shows the time-series profiles of the training data-set.

|  | household_power | solar_power | wind_power | EUR/kWh | renewable_power | House&RW_load | total_cost |
| --- | --- | --- | --- | --- | --- | --- | --- |
| time |  |  |  |  |  |  |  |
| 2018-01-01 00:00:00 | 4.006786 | 0.0 | 7.644 | -0.00527 | 7.644 | -3.637214 | 0.004792 |
| 2018-01-01 00:15:00 | 3.886011 | 0.0 | 7.644 | -0.00527 | 7.644 | -3.757989 | 0.004951 |
| 2018-01-01 00:30:00 | 3.833506 | 0.0 | 7.644 | -0.00527 | 7.644 | -3.810494 | 0.005020 |
| 2018-01-01 00:45:00 | 4.332042 | 0.0 | 7.644 | -0.00527 | 7.644 | -3.311958 | 0.004364 |
| 2018-01-01 01:00:00 | 4.034777 | 0.0 | 8.064 | -0.02999 | 8.064 | -4.029223 | 0.030209 |

Figure 3.6: Example of a VPP input data-set table (validating data-set).

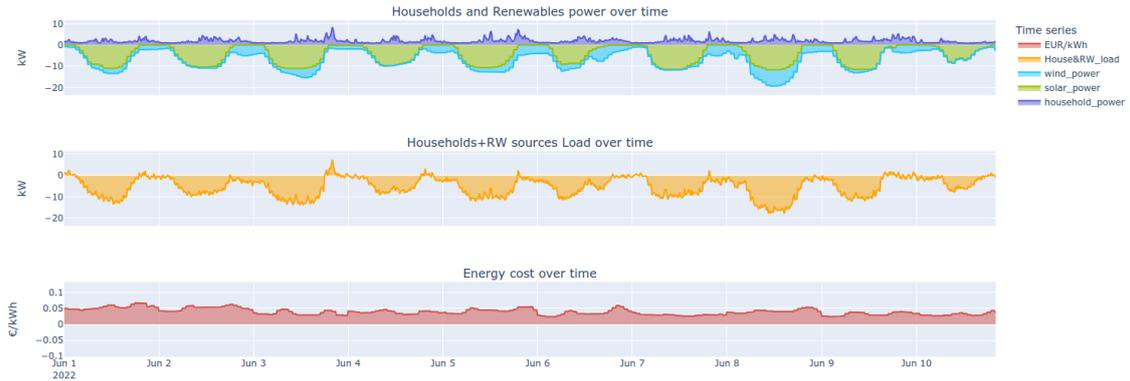

Figure 3.7: Visualization of the training data-set.

**Data-set goal**

Each data-set has its own values for the total energy consumed, produced, and cost. These summarizing values depend on the stochastic behaviour of the sampled time-series of renewable power generation, household power consumption, and energy market prices. According to these values unique for each data-set, it is possible to evaluate a **Maximum average EVs available energy at departure**.

This parameter represents the energy that every EV of the episode would reach if all the unused energy produced by the Renewable sources (not consumed by the households) was entirely equally stored in the EVs, having the defined average SoC at arrival (50% in this research). It is evaluated through the formula:

$$Av\_EVs\_en_{max} = Av\_EVs\_en_{arriving} + \frac{(renewable\_en_{supply} - households\_en_{demand})}{tot\_charging\_events\_N} \quad (3.5)$$

Where $Av\_EVs\_en_{arriving}$ is the average energy level of EVs at arrival (evaluated as $Av\_EVs\_SoC_{arriving} \cdot EV\_capacity$), the $renewable\_en_{supply}$ is the total energy produced by the renewable sources, the $households\_en_{demand}$ is the total households power demand and the $tot\_charging\_events\_N$ is the total number of charging events in the simulation.

The maximum average EVs available energy is thought to be **unreachable** because its feasibility is based on the assumption that there is zero grid import/export and the EVs are always present. Even if these assumptions would be satisfied, this value is still improbable to reach because of the aleatory imperfections of the simulation, like the limited storage capacity of the EVs. This parameter is therefore defined as **unfeasible** to reach. However, it is still a meaningful parameter that describes the data-set limits and it is considered the ultimate goal of the Virtual Power Plant agent. Therefore, the maximum average EVs available energy will be used to compare the VPP results. This value is calculated and stored during the initialization of the environment, and it is used along the episodes to compare the obtained results.

# Chapter 4

# Methodology

This chapter describes the final environment shape and the definition process that took place during the research according to the Reinforcement learning agent structure and developing features. It finalizes the solution proposed for the Virtual Power Plant of EVs problem.

## 4.1  RL framework update

The tests performed on the first development of the VPP RL framework led to poor results and suggested an update of the framework itself. The first RL algorithms were tuned and trained in the previous environment for some millions of steps (each training lasted from 1 to 7 or 8 hours) and the training results were not demonstrating a progressive policy optimization. Therefore, an updated version of the RL framework was designed, as it is shown in Figure 4.1.

**The Environment states' spaces**

The first assumption, based on the fact that the agent needed significant state data as much as possible, resulted to be wrong. The observation space was too wide for the agent to be able to recognize patterns and define an effective policy. The observation space included five different continuous state variables, two of which were arrays of multiple continuous variables. Therefore, the observation space was restricted to only three variables, one of which is an array. The new observation space variables can be described as follows:

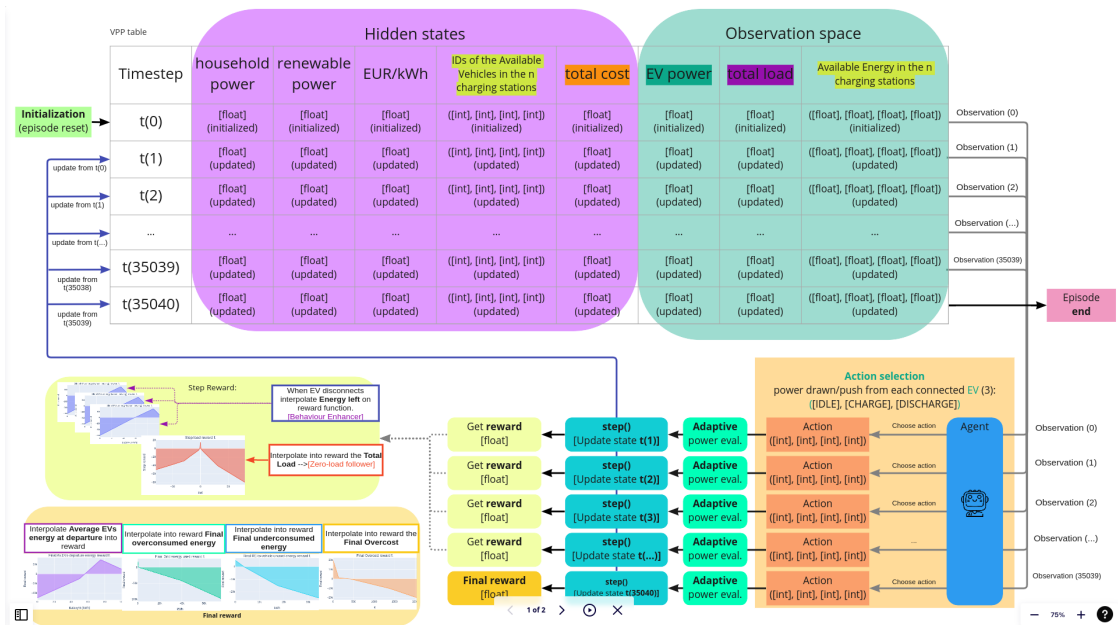

Figure 4.1: The updated idealization of the VPP RL framework structure.

- EV power: a Box type (continuous) scalar indicating the total power drawn/pushed from the EVs at time-step $t$.

  - It is evaluated by summing the actual power drawn/pushed from/to the connected EVs.

  - It ranges from the maximum charging station power defined in the Elvis configuration file and its negative value.

  - It depends on the actions chosen by the agent, and by the available energies of the EVs connected to the charging stations state.

- Total load: a Box type (continuous) scalar indicating the resulting power from the supply/demand sum at time-step $t$.

  - It is evaluated every time-step $t$ by the formula (converting to kWh, a time-step $\delta t$ is 15 minutes or $\frac{hour}{4}$ expressed in hours):

$$\text{total\_cost}_t[EUR] = \text{total\_load}_t[kW] \cdot \delta t \text{ [hours]} \cdot \text{price\_energy}_t[EUR/kWh]$$

- It ranges from the maximum household consumption plus the maximum charging station power defined in the Elvis configuration file and its negative value. It is expressed in Euros (EUR).

- Available energies: an array of $n$ Box type (continuous) scalars of the available energies at the charging stations, $n$ being the number of charging stations in the simulation.

    - Each array value ranges from 0 (no EVs connected) to 100 kWh (the maximum EVs capacity). Minimum feasible charge if an EV is connected is 0.1 kWh.

    - It depends on the Elvis simulation charging events list and on the actions taken in the past (charging/discharging connected EVs battery).

This arrangement allows more hidden state variables in the VPP simulation that are not fed to the agent in the observation space: the household power, the renewable power, and the energy market prices. These time-series come from the input data-set (different series for training, testing, and validating data-set) and are not directly accessible by the agent, but they are indirectly extractable from the observation space variables. The agent shall learn the patterns of these hidden variables by correlation with the *total_load* and *total_cost* spaces.

**Action space update**

The action space in the previous configuration was too wide. The agent could potentially choose an immense set of actions for each time-step. This immense choice was not favourable for the exploration of the best policy but was instead leading to inconclusive local minima. The local minimum usually found was the constant idle action for all charging stations, which corresponds to not doing anything (charging/discharging) during the whole simulation. The updated idealization of the action space consisted of a Multi-Discrete space with 3 actions per charging station. Each action per charging station, numbered from 0 to 2, has its own power transfer direction and intensity:

- Action 0: Idle. No power drawn/pushed from the EV.
- Action 1: Charge [from 1 kW to 11 kW for $\delta t$] the connected EV.
- Action 2: Discharge [from -1 kW to -11 kW for $\delta t$] the connected EV.

**Adaptive power evaluation**

With the action set defined in the previous section, the power drawn/pushed from the vehicle ranges from the minimum to the maximum value of the charging station power. The problem with the discrete power selection of the previous implementation was that it was **imprecise** and the zero-load supply/demand balance was never perfectly met, and hardly generally met. Overall, the discrete negative/positive power selection resulted poorly in the agents learning any action set that was not the Idle action for all the charging stations for the whole episode. Therefore, the exact power needed from the infrastructure must be evaluated through an adaptive power evaluation. This adaptive power simply consists in getting the power value necessary to balance the supply/demand power and to reach a zero total load. However, the direction of the power flow in the charging points is still decided by the RL agent. The power flow is always comprised between the limits selected in the Elvis configuration file: 1 kW (minimum charging station power) and 11 kW (maximum charging station power). The rated power of the charging stations is 3.7 kW. We distinguish four main cases of the adaptive power behaviour (for a selected action different from Idle in a charging station with a connected EV):

- Households and Renewables load > 0 kW and Discharge action selected: Negative power selected from the EV between -1kW and -11kW that compensates the supply/demand load.

- Households and Renewables load > 0 kW and Charge action selected: Positive power selected from the EV equal to the rated power charging station (3.7 kW).

- Households and Renewables load < 0 kW and Discharge action selected: Negative power selected from the EV equal to the rated power charging station (-3.7 kW).

- Households and Renewables load < 0 kW and Charge action selected: Positive power selected from the EV between 1kW and 11kW that compensates the supply/demand load.

**Actions validity**

The EV load state variable is evaluated by summing the power chosen by the agent for each station. The Agent must pick an action for each charging station. The actions' effects are subject to some restrictions:

- Actions different from 0 (Idle) taken for a charging station at time-step $t$ without an EV connected are useless and do not change any state.

- Charging an EV with a resultant energy greater than 99.9 kWh does not produce any effect (the excess energy over the 99.9 kWh is reconverted in power and subtracted from the EV power).

- Discharging an EV with a resultant energy smaller than 0.1 kWh does not produce any effect (the excess energy surpassing the minimum threshold of 0.1 kWh is reconverted in power and subtracted from the EV power).

- EVs connected with a SoC below 10% (10 kWh in this environment) cannot be discharged or kept idle. They will be charged instead with Action 1 if Action 3, or Action 5 were not selected for that EV.

- EVs connected with a SoC below 20% (20 kWh in this environment) cannot be discharged. They will be kept Idle instead if Action 1, Action 3, or Action 5 were not selected for that EV.

These rules are the same that will asses whether the action taken was valid or invalid. The actions' validity concept is used in the Action Masking learning approach explained in the next chapters. It is possible to identify each possible action in a numbered table as shown in Figure 4.2. In this visualization, actions can be numbered from 0 to a maximum value of 11, obtained with the same equations described in the analysis chapter. Since for each time-step the agent must pick an action for each charging station (4 in the studied environment), the action set is composed of 81 action combinations ($3^4$).

| Action codes table | | Charging station $n$ | | | |
|---|---|---|---|---|---|
| Action meaning | Action Value | 0 | 1 | 2 | 3 |
| Idle [+0 kW] | 0 | 0 | 1 | 2 | 3 |
| Charge [+1, +11 kW] | 1 | 4 | 5 | 6 | 7 |
| Discharge [-1, -11 kW] | 2 | 8 | 9 | 10 | 11 |

Figure 4.2: The updated table of the VPP action space.

**Updated Reward functions**

The VPP simulation evaluates a reward for the observation space for each time-step as explained in the analysis chapter. An additional final reward at the last step of the simulation is evaluated even in the final VPP environment version. The reward functions applied during the simulations are 6 in total, 2 being applied at every step and 4 being applied at the end of the simulation. A focus from the RL framework on the reward functions is available in Figure 4.3.

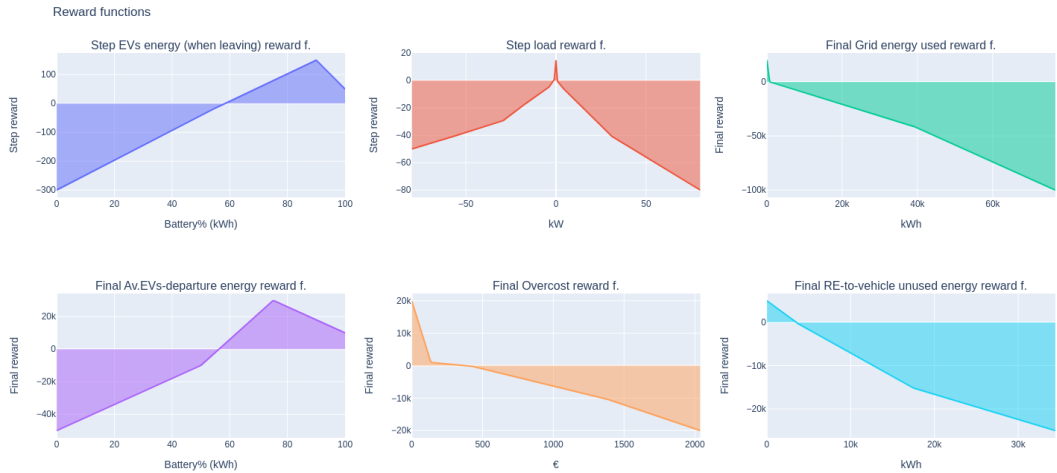

Figure 4.3: Final reward functions implemented in the VPP environment.

In the final idealization of the VPP RL framework, the reward for each time-step is evaluated considering only 2 observation state variables:

- The reward attributed to the total load value at time-step $t$: The total load value is mapped into an interpolated reward function peaked at 0 kW, positive between -1.5 kW and 1 kW and negative otherwise. This is the main component with the higher weight on the step and total reward that enables the VPP agent to work as a zero-load follower.

- The reward attributed to the available energy at the departure of each EV. This reward is evaluated exclusively when an EV leaves the charging station, and its available energy percentage value is mapped into an interpolated reward function positive after 55% and peaked at 90%. This reward is given only when any EV disconnects from the charging station, and it therefore acts as "Behaviour enhancer". It will reward a good job resulting in a high percentage of energy left, and it will punish a low percentage of energy left.

The final reward evaluated at the end of the simulation is instead divided as follows:

- The reward attributed to the Average EVs available energy at departure: the average energy percentage value is mapped into an interpolated reward function positive after 55% and peaked at the maximum average EV available energy, a parameter described in the previous chapter.

- The reward attributed to the overall grid energy used (grid import): The summed load values drawn from the grid, converted to energy are mapped into an interpolated reward function peaked at 0 kWh, positive between 0 kWh and 10% of the grid energy used in the uncontrolled Elvis simulation, and negative otherwise.

- The reward attributed to overall RE-to-vehicle (RE2V) unused energy (grid export): the sum of the unused load from the renewable sources, transformed to energy, is mapped into an interpolated reward function peaked at 0 kW, positive between 0 kWh and 20% of the RE-to-vehicle (RE2V) unused energy of the uncontrolled Elvis simulation, and negative otherwise.

- The reward attributed to the overall grid energy cost: The summed load cost values drawn from the grid are mapped into an interpolated reward function peaked at 0 kWh, positive between 0 kWh and 10% of the Grid energy cost of the uncontrolled Elvis simulation, and negative otherwise.

**Rewards statistics**

The VPP environment embeds functions to display the reward obtained over time during each step as it can be seen in Figure 4.4. The plots in Figure 4.4 show the TRPO algorithm's reward for one month of VPP simulation:

- The first series shows the reward obtained for the total load of each time-step, overlapping with the total load profile.

- The second series shows the reward obtained for the EVs departure-energy in red, and the total reward in green (load reward plus EVs departure-energy reward), overlapping with the number of available EVs over time.

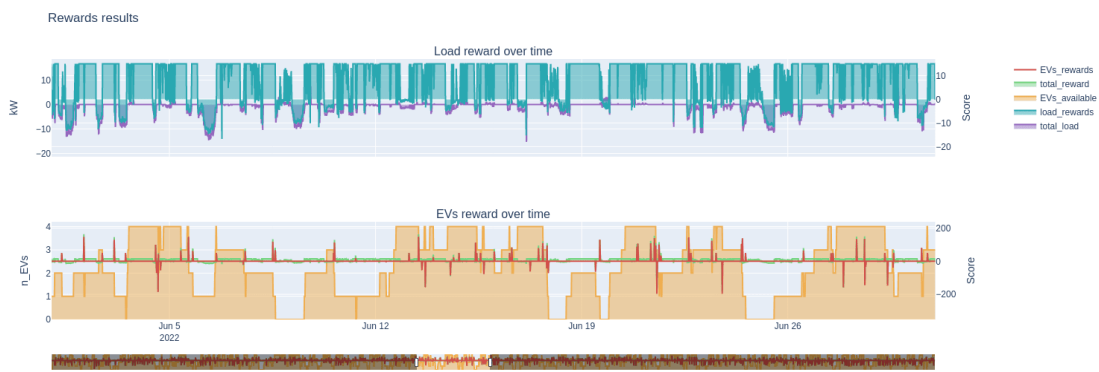

Figure 4.4: TRPO reward over time plot

The VPP environment can also display the components of the cumulative reward. The cumulative reward can be broken down in the step reward sources and the final rewards as shown in Figure 4.5.

## 4.2 Updated simulation key-parameters

**Simulation key-parameters**

The main index of performance of an algorithm or a VPP simulation is the **cumulative reward**, obtained by summing the step rewards and the final rewards analyzed in the previous section. Moreover, to acquire a more detailed view of a simulation's results, the cumulative reward can be broken down into the key-parameters that had been interpolated to obtain it. The names of these parameters are:

- Grid energy used (grid import): the energy drawn directly from the grid infrastructure that must be paid. Expressed in kilo Watt-hours, also referred to in the next chapters as "over-consume".

- RE-to-vehicle unused energy (grid export): the energy produced by the renewable sources that is not consumed by the households nor stored in the EV batteries. Expressed in kilo Watt-hours, also referred to in the next chapters as "under-consume".

- Grid energy cost: the cost of the energy exclusively drawn from the grid infrastructure. Expressed in Euros, also referred in the next chapters as "over-cost".

- The average EVs available energy at departure: it is the average of the available energy in the EVs batteries when leaving the charging stations, for the whole simulation. Expressed in kilo Watt-hours, is contained between 0 and 100 kWh, and it is also referred to in the next chapters as "av_EV_energy_left". In an uncontrolled charging Elvis simulation, this value reaches easily 100 kWh, while in a VPP controlled charging simulation it is to be compared with the corresponding maximum average EVs available energy at departure. The latter is described in the previous chapter, also referred to as $Av\_EVs\_en_{max}$.

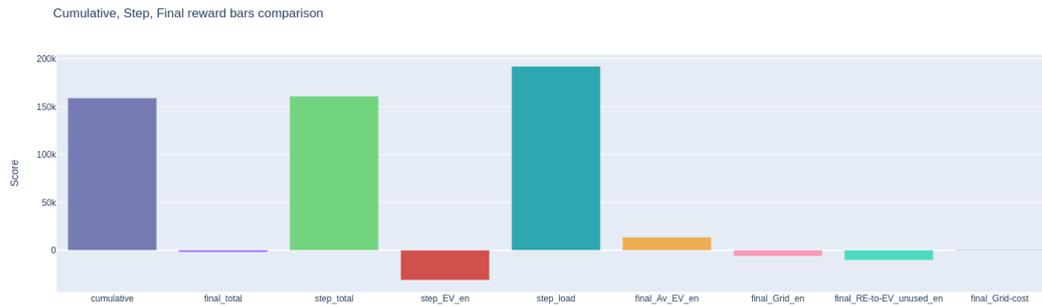

Figure 4.5: A2C algorithms reward statistics results

These parameters are the most decisive for the evaluation of the VPP agent because they summarize how many good actions the agent took during the simulation. The highest priority is given to the grid energy used since the scope of the research is trying to design an auto-sufficient MG with access to the grid as close as possible to zero. The strict correlation between the grid energy cost and the grid energy used gives a double weight to this parameter. Then, the RE2V unused energy is considered the second main parameter of importance as we want to optimize the MG as much as possible by exploiting all the available energy. The RE2V unused energy (grid export) is also strictly correlated to how much energy is stored in the EVs. The other parameters presented are:

- Households and Renewable energy summed: considering solely VPP input data-set load time-series, it is the sum of all the energy values obtained during the simulation (energy evaluated for each time-step $t$ as:

$$energy\_sum_t = total\_load\_sum_t \cdot \delta t \qquad (4.1)$$

  with $total\_load\_sum_t$ being the sum of the total load time-series. Expressed in kilo Watt-hours. It considers the negative energy as energy sold back to the grid (out of the scope of the thesis).

- Supply/demand energy difference: In an Elvis uncontrolled charging simulation, it is the sum of all the energy values obtained during the simulation (energy evaluated for each timestep $t$ as

$$energy\_sum_t = total\_load\_sum_t \cdot \delta t \qquad (4.2)$$

  Expressed in kilo Watt-hours. It considers the negative energy as energy sold back to the grid (out of the scope of the thesis).

- Total cost (if selling excess energy): the sum of all the cost values obtained during the simulation. Expressed in Euros. It considers the negative energy as energy sold back to the grid (out of the scope of the thesis).

- Grid energy cost: The sum of all the costs of the energy exclusively drawn from the grid during the whole simulation. It must be greater or equal than zero.

| Simulation type | Parameter | Training | Testing | Validating |
|---|---|---|---|---|
| Plain data-set | Households and Renewable energy summed | (produced) -34117.7 kWh | -21214.64 kWh | -30085.39 kWh |
| | Grid energy used | 1556.25 kWh | 4947.18 kWh | 2136.67 kWh |
| | (RE2V) RE-to-vehicle unused energy | 35673.95 kWh | -26161.81 kWh | -32222.06 kWh |
| | Total cost (if selling excess energy) | (earned) -1196.64 € | -489.75 € | -1187.15 € |
| | Grid energy cost | 97.86 € | 233.11 € | 113.34 € |
| Elvis uncontrolled | Supply/demand energy difference | 8217.79 kWh | 21434.65 kWh | 13003.47 kWh |
| | Grid energy used | 31697.31 kWh | 38651.23 kWh | 34535.56 kWh |
| | (RE2V) RE-to-vehicle unused energy | 23479.51 kWh | 17216.57 kWh | 21532.09 kWh |
| | Total cost (if selling excess energy) | (spent) 454.65 € | 887.56 € | 619.13 € |
| | Grid energy cost | 820.03 € | 1373.29 € | 1502.09 € |
| | Charging events | 1043 | 1043 | 1043 |
| Optimal VPP goal | Grid energy used | 0 kWh | 0 kWh | 0 kWh |
| | (RE2V) RE-to-vehicle unused energy | 0 kWh | 0 kWh | 0 kWh |
| | Grid energy cost | 0 € | 0 € | 0 € |
| | Average EVs available energy at departure | 84.2 kWh | 75.08 kWh | 80.89 kWh |

Figure 4.6: Key-parameters' results table per data-set

The table of key-results extrapolated from every simulation from the data-sets is presented in Figure 4.6. These results are obtained by initializing the environment with the same Elvis configuration file having the average EV SoC set to 50.0 % and the weekly number of EV arrivals set to 20. Three sets of results are presented for each data-set:

- The first set of results regards a simulation without any EVs integration, which can be also considered a data-set description to understand the produced and consumed energy limits.

- The second set of results is obtained by integrating the uncontrolled EV charging infrastructure from Elvis. These values will be used as a reference to compare the VPP agent results with controlled charging.

- The last set of results simply indicates the optimal values the VPP agent should try to reach for the key parameters described above. It includes also the optimal average EVs available energy parameter described in the previous chapter.

**Data-sets self-consumption and autarky**

The key parameters described in the previous section are used also in the evaluation of two other meaningful parameters that will be used to assess the VPP performances:

- The self-consumption rate: the ratio between the renewable energy produced and the portion of the renewable production self-consumed by the system, with 0% renewable self-consumption meaning that all the produced energy is exported to the grid. 100% of self-consumption means that all the renewable energy produced is directly consumed or stored.

- Autarky rate: the ratio between the whole energy consumed from any source, and the energy consumed coming from the renewable energy produced. The energy consumed coming from renewable sources includes the energy directly consumed when produced and the energy stored and then later used.

These parameters are famous in the field of auto-sufficient energetic systems and Net Zero-Energy Buildings (NZEB) for their relevance in describing the supply/demand balance of systems for time periods [36]. The VPP simulations will produce self-consumption and autarky rates considering also the renewable power stored in the EVs, but the plain input data-sets can already express those rates by means of direct consumption as shown in Figure 4.7.

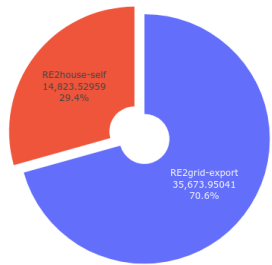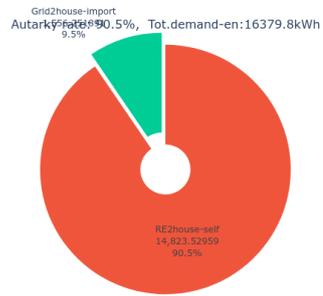

(a) Training data-set.

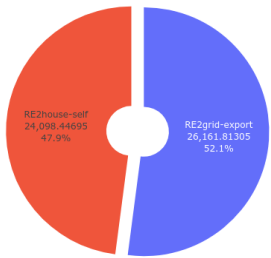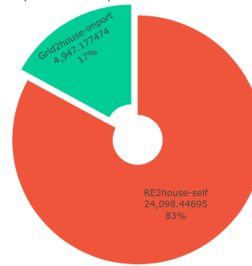

(b) Testing data-set.

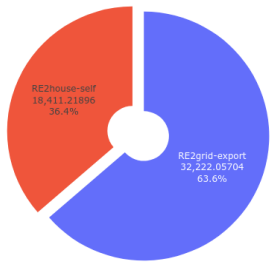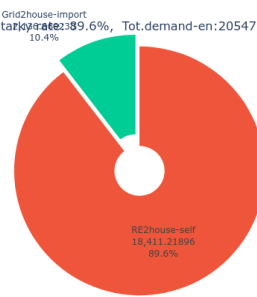

(c) Validating data-set

Figure 4.7: Self-consumption and Autarky rate charts for each input data-sets.

# Chapter 5

# Algorithms implementation and comparison

This chapter dives into the main task of the thesis: the implementation of reinforcement learning algorithms for the virtual power plant environment. In detail, the training algorithms strategy, processes, outcomes, and performances. It concludes with the proposal of the best-suited algorithm for the solution implementation.

## 5.1 The Hyper-parameter tuning

All the RL algorithm models of this research are taken from the Stable-Baselines3 (SB3) library. The proposed algorithms are the only ones satisfying the environment space features described in the Analysis chapter: a dictionary of state variables space and a Multi-Discrete action space. For each algorithm proposed, the Hyper-parameters tuning process is performed through the Weights & Biases (W&B) Sweep tool. The parameters to be tuned are set in the algorithm sweep configuration file. The WB Sweep consists of a pipe-lined series of 500000-steps models training with certain hyper-parameter sets run on the **testing data-set**. The hyper-parameters research approach of this thesis is configured to a **random search with early stopping**, which compared to the Grid and Bayesian search proved to be much more effective as explained in the W&B Hyperband paper [37]. The hyper-parameter set varies for each algorithm, but since they are all policy gradient algorithms, there are many parameters in common:

- Policy type: the policy that the algorithm will optimize. For this research, all the algorithms use the Multi-input Policy that is designed for Multi-discrete actions environment.

- Batch size: the mini-batch training step size. Since a one-year simulation is composed of 35041 steps (from step 0 to 35040), one month or one-twelfth of the simulation is 2920 steps; therefore the batch size considered values were multiples of 2920 up to 17520 steps (six months, half simulation).

- Number of steps (n_steps): The number of steps to run for each environment per update (i.e. batch size is the number of steps multiplied by the number of environment copies running in parallel).

- Learning rate: it can be a function of the current progress remaining, but setting a constant value proved to be more effective since the policy optimizer already modify the learning rate value as the simulation goes on. Ranging from 1 to 0, it was uniformly set between 0.0001 and 0.0014.

- Total time-steps: the training number of time-steps or the total number of samples (environment steps) to train on. Set to 500000 for every algorithm's Hyper-parameter sweep (approximately 15 epochs or 15 years of simulation).

- Gamma: the discount factor to reduce the value of the future state relevance since we want to emphasize more on the current state rather than the future state. Logarithmic uniform selection between 0.9999 and 0.89.

- Generalized advantage estimation lambda (gae_lambda): Factor for the trade-off between bias and variance for the Generalized Advantage Estimator [38]. Selected between 0.8 and 1.

- Clip range: the clipping parameter. Uniformly selected between 0.1 and 0.4.

- Maximum gradient clipping norm (max_grad_norm): The maximum value for the gradient clipping [39]. Uniformly selected between 0.1 and 5.

- Entropy coefficient (ent_coef): the Entropy coefficient for the loss calculation. the entropy determines how random the decisions of the model are. Logarithmic uniform selection between 1e-10 and 0.1.

- Value Function coefficient (vf_coef): the value function coefficient for the loss calculation. It constitutes, together with the entropy loss, another loss term of the surrogate loss function in the objective function. [40]. Uniformly selected between 0.0 and 1.0.

- The policy keyword arguments:
  - Orthogonal initialization (ortho_init): whether to use or not the orthogonal initialization. This parameter was set to true for all algorithms as it proved to give better results.

- Activation function (activation_fn): The activation function module loaded from Pytorch classes. Chosen among the most successful "Tanh" and "ReLU". The Hyperbolic tangent "Tanh" proved to give better results.
- The Optimizer class: chosen among Adam, Stochastic Gradient Descent (SGD), and RMSprop (root mean square propagation). The RMSprop optimizer [41] proved to lead to better results in the alike PPO algorithms.
- Network architecture (net_arch): The specification of the policy and value networks, i.e. the neural network layers for the value and best action prediction as explained in the SB3 custom network definition guide [42]. Selected among different proposed sizes and combinations of shared/separate networks for the value and action prediction.

The Hyper-parameter Sweep will highlight the best configuration by comparing the simulations' cumulative rewards as the main index of performance. The parameter set coming out from the Sweep that leads to the highest cumulative reward is then trained once again in a new model for 1500000 steps on the **training data-set**. Additionally, the summarizing results used to further analyze and compare the algorithms are the key output parameters described in the Methodology chapter: the Grid energy used (also referred in the graphs as "over-consume"), the RE-to-vehicle (RE2V) unused energy (also referred to in the graphs as "under-consume"), the Grid energy cost (also referred in the graphs as "over-cost") and the average EVs available energy at departure (also referred in the graphs as av_EV_energy_left). Particular importance is given to the RE-to-vehicle (RE2V) unused energy (grid export), the average EVs available energy at departure, and the Grid energy used (grid import) because they already describe the VPP simulation performance by plotting these parameters in a three-dimensional space (respectively, on the x,y,z axes).

## 5.2 Advantage Actor-Critic algorithm (A2C)

The Advantage Actor-Critic model is the simplest model among the ones available in the SB3 that satisfies the environment space features, as it was described in the Background theory chapter. It was also the first one implemented in the VPP environment.

### 5.2.1 Hyper-parameters tuning

The A2C algorithm was tuned according to the parameters in the table represented in Figure 5.7, which shows for each line the cumulative reward obtained for each hyper-parameter set. The results for the key output parameters are summarized in the line graph represented in Figure 5.1a, where a certain linearity can be already seen. The connecting lines suggest a direct proportionality between the Grid energy used (overconsume), the Grid energy cost (overcost), and the average EVs available energy at departure (av_EV_energy_left), as consuming more extra energy from the grid leads to higher costs and higher energy stored in the EVs. A negative proportionality can be seen instead between the RE2V unused energy (underconsume) and the average EVs available energy at departure (av_EV_energy_left), since less wasted energy results in more energy stored in the EVs. The 3D-plot of the summarizing key parameters of Figure 5.1b (the color scale of the scattered points is the third dimension) confirms such negative linearity, suggesting that lower RE2V unused energy values (in the x-axis) lead to higher average EVs available energy at departure values. The cumulative rewards range between very negative results down to -150000 points, and not sufficient results up to 40000 reward points.

### 5.2.2 Training results

The best hyper-parameters set corresponding to the highest score was further trained in a new model for 1500000 steps on the **training data-set**. This section presents the results and performances along the VPP simulation of this trained model, run on the testing data-set. Figure 5.2 shows the available energies per charging station during one month of the VPP simulation controlled by the A2C agent. The available energy lines tend to either quickly raise to 100 kWh or quickly descend to 20 kWh, which is the minimum allowed by the agent. This behaviour reveals that the agent did not learn to follow the trends of the data-set, but rather learned to attribute some EVs as charging devices and others as discharging ones. This behaviour is confirmed by the histogram of the EVs' available energy at departure distribution available in Figure 5.8c.

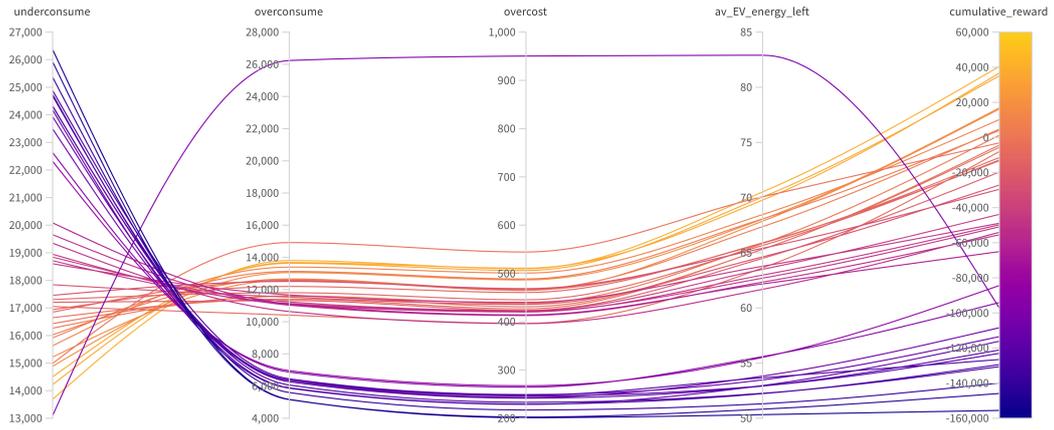

(a) A2C hyper-parameters sweep line results

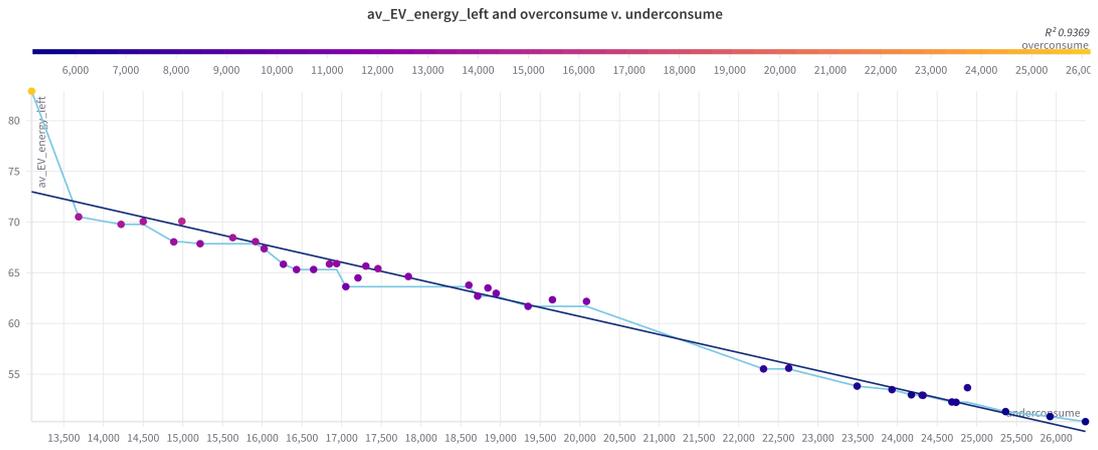

(b) A2C algorithm-sweep 3D-plot of the summarizing key results

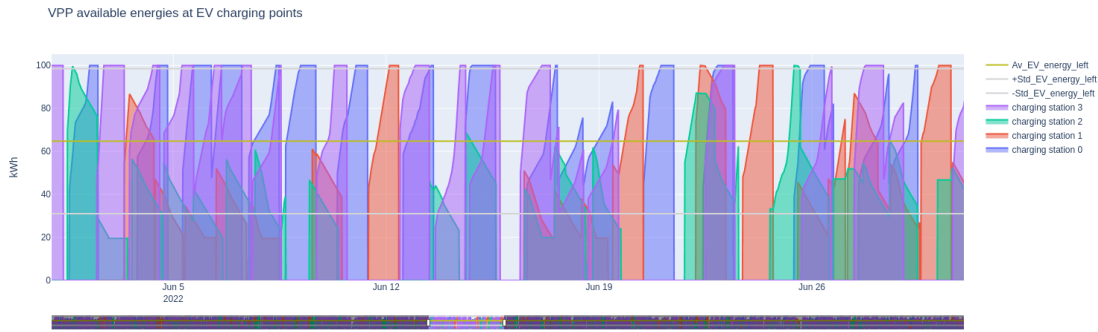

Figure 5.2: A2C VPP available energies per charging station over one month.

## 5.3 Maskable Proximal Policy Optimization algorithm (MaskablePPO)

The MaskablePPO algorithm has the same structure and operation as a PPO algorithm as explained in the background chapter, but it also integrates the action masking feature. This algorithm was potentially considered the best algorithm for policy optimization of the VPP environment. The action masking feature is indeed a promising learning advantage because it creates a mask of "invalid actions". This mask is based on useful actions for the VPP goal, for each time-step of the simulation. The invalid actions are evaluated according to the conditions described in the analysis chapter. A heat-map of the valid/invalid actions taken by the MaskablePPO trained agent is shown in Figure 5.3. The valid-actions heat-map presented on the left reveals how many valid actions were taken by the agent (yellow zones) for each station (each station is assigned to one of the four columns). The vehicle-availability heat-map on the right completes the information by highlighting when no vehicle is connected to the charging stations during the simulation (the dark blue zones, for each column).

The evaluated mask modifies the logits corresponding to the invalid actions with negative infinity before passing the logits to softmax, thus setting to a virtual zero the gradient towards the direction of those actions. The algorithm was considered promising because in the first implementation of the VPP environment described in the analysis chapter the action space was too wide, and the action masking helps the agent move in such a wide space. For this reason, the A2C and the PPO algorithm were not able to perform a sufficient action exploration and they easily got stuck in a local minimum of the action space (usually the IDLE action for all stations [0,0,0,0]). The MaskablePPO algorithm performed better because instead of learning the correct actions by getting punishments (negative rewards) for invalid actions, the agent would have more probability of choosing correct actions by default, speeding up the learning process. The MaskablePPO algorithm allowed the agent to learn how to minimize the grid energy used (grid export). However, the agent would still choose invalid actions and the results were still unsatisfactory because the VPP agent did not learn how to minimize the RE2vehicle unused energy (grid export).

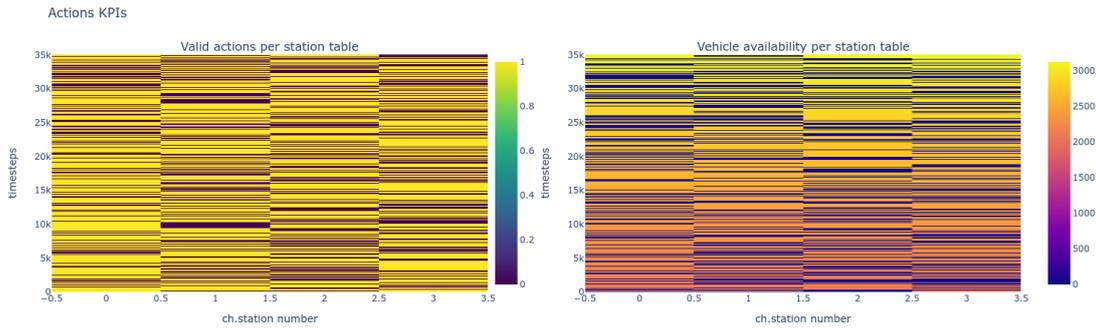

Figure 5.3: MaskablePPO actions mask and available EVs at charging stations Heatmaps.

### 5.3.1 Hyper-parameters tuning

The results of the MaskablePPO algorithm hyper-parameter tuning are presented in Figure 5.5. The results reach higher levels of cumulative reward, but they are still considered unsatisfying.

### 5.3.2 Training results

Even the best hyper-parameters set corresponding to the highest score, which was further trained in a new model for 1500000 steps on the **training data-set** performed poorly. The behaviour revealed by the histogram of the EVs' available energy at departure shown in Figure 5.8d suggests that the agent did not charge the EVs enough, maintaining the concentration of the distribution at medium levels of energy. The best MaskablePPO hyper-parameters set is shown in the table represented in Figure 5.7.

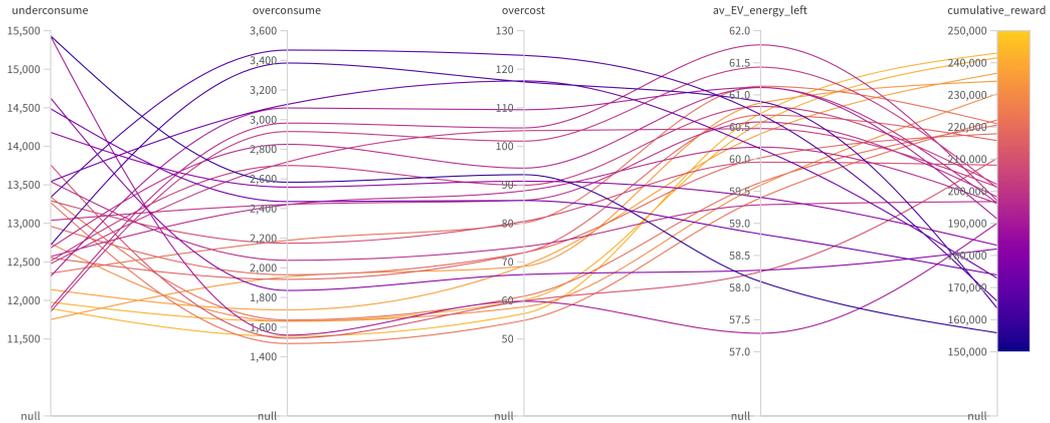

Figure 5.4: MaskablePPO algorithm hyper-parameter tuning results.

## 5.4 Trust Region Policy Optimization algorithm (TRPO)

The TRPO algorithm was the first algorithm that started producing good results. The TRPO algorithm was not considered promising because the trust region method described in the background chapter can potentially lead to a local minima trap. TRPO is also generally considered an over-complicated algorithm, considering its successful first-order implementation PPO. However, the trust region method for the policy optimization proved to succeed in the action exploration and the agent managed to learn how to maximize the cumulative reward.

### 5.4.1 Hyper-parameters tuning

The results of the TRPO algorithm hyper-parameter tuning are presented in Figure 5.5. The episodes reach high levels of cumulative reward and generally good results.

### 5.4.2 Training results

The best hyper-parameter set corresponding to the highest score, further trained in a new model for 1500000 steps on the **training data-set**, performed very well. The behaviour revealed by the histogram of the EVs' available energy at departure shown in Figure 5.8e shows how the distribution increases its concentration for high levels of energy. This distribution shape translates to a good value of Average

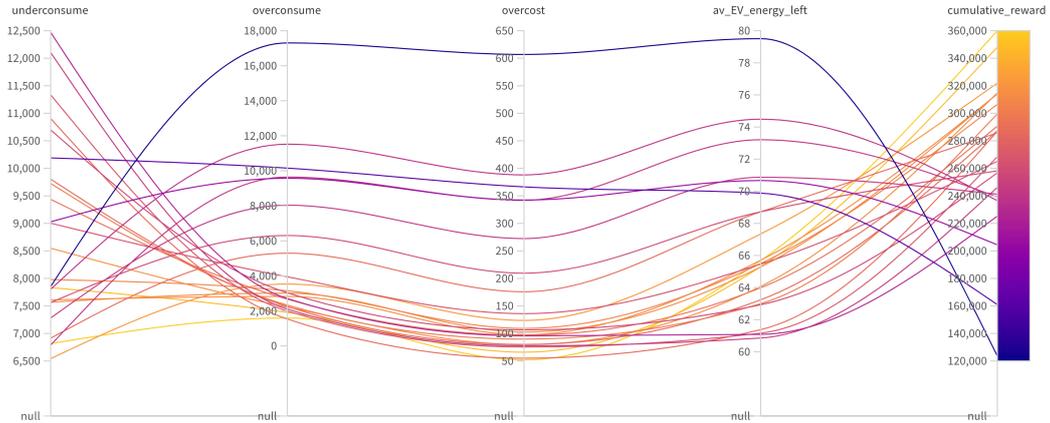

Figure 5.5: TRPO algorithm hyper-parameter tuning results.

EVs energy left at departure, coupled with very low levels of Grid used energy (grid import) and Unused RE2vehicle energy (grid export). The best TRPO hyper-parameters set is shown in the table represented in Figure 5.7.

## 5.5 Recurrent Proximal Policy Optimization algorithm (Recurrent PPO)

The RecurrentPPO algorithm proved to be the best for this research. The RecurrentPPO algorithm, also known as LSTM (Long-Short-Term Memory) PPO, is based on the policy optimization that supports "hidden states" observation. For each time-step, the model would generate an LSTM array with weighted values connected to the detected hidden states. The action prediction is then based on this array. This algorithm's feature perfectly fits the VPP environment updated framework where the observation space includes only a few state variables, and the others are left "hidden" as described in the methodology chapter. The RecurrentPPO proved to succeed in the action exploration and the agent managed to learn how to maximize the cumulative reward.

### 5.5.1 Hyper-parameters tuning

The results of the RecurrentPPO algorithm hyper-parameter tuning are presented in Figure 5.6. The episodes reach high levels of cumulative reward and generally good results.

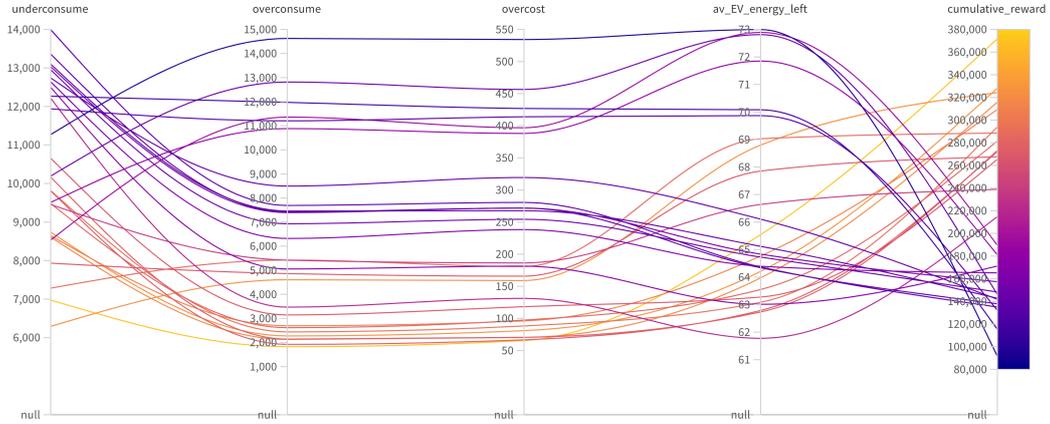

Figure 5.6: RecurrentPPO algorithm hyper-parameters sweep results.

### 5.5.2 Training results

The best hyper-parameter set corresponding to the highest score, further trained in a new model for 1500000 steps on the **training data-set**, performed better than any other algorithm. The behaviour revealed by the histogram of the EVs' available energy at departure shown in Figure 5.8f is similar to the TRPO distribution, but with less concentration at very low levels of energy left, and higher ones at 100%. The RecurrentPPO's results include the highest Average EVs energy left at departure, low levels of Grid used energy (grid import), and the lowest levels of Unused RE2vehicle energy (grid export). The best RecurrentPPO hyper-parameters set is shown in the table represented in Figure 5.7.

## 5.6 Algorithms comparison charts

This section presents the comparison between the best results per algorithm. The hyper-parameter sets chosen for each algorithm are shown in the table in Figure 5.7, together with the cumulative reward results per algorithm for each data-set. Figure 5.8 presents the EV departure-energy histograms per algorithm. The histograms for an Elvis uncontrolled simulation and a VPP simulation with random actions are also presented for comparison. Figure 5.9 shows the VPP simulation key-parameters bar-chart for the trained algorithms compared with the Elvis uncontrolled simulation. The bar values of Figure 5.9a can be found in the table presented in Figure 5.9b.

| Hyper-parameters results table | Algorithms | | | |
|---|---|---|---|---|
| Parameter | A2C | MaskablePPO | TRPO | RecurrentPPO |
| n_steps | 8760 | 11680 | 17520 | 8760 |
| batch_size | 8760 | 11680 | 17520 | 8760 |
| total_timesteps | 1500000 | 1500000 | 1500000 | 1500000 |
| learning_rate | 0.000714503095438 | 0.001333352575388 | 0.000654435637725 | 0.001033322426833 |
| clip_range | - | 0.4 | - | 0.2 |
| gamma | 0.915907895302168 | 0.96858167756061 | 0.895016629337843 | 0.951348296198191 |
| gae_lambda | 0.8 | 0.92 | 0.9 | 0.92 |
| ent_coef | 1.50E-07 | 0.027311787524163 | 0 | 2.09E-09 |
| vf_coef | 0.011059086790669 | 0.942525617371529 | | 0.132911682993952 |
| max_grad_norm | 0.7 | 5 | 0 | 5 |
| line_search_max_iter | - | - | 10 | - |
| line_search_shrink_factor | - | - | 0.6 | - |
| optimizer class | Adam | RMSprop | RMSprop | Adam |
| net_arch | medium | Small-separate | big | Medium-short |
| Training cumulative_reward | 140939.05 | 222022.25 | 347817.53 | 391583.41 |
| Testing cumulative_reward | 159089.05 | 243022.28 | 355838.29 | 383539.9 |
| Validating cumulative_reward | 166890.68 | 231422.11 | 360786.87 | 396772.12 |

Figure 5.7: Algorithms hyper-parameters best sets table.

## 5.7 Best algorithm implementation: RecurrentPPO

The results in the table of Figure 5.7 highlighted the RecurrentPPO algorithm as the best one since the RecurrentPPO model reached cumulative reward scores higher than 380000 in the testing data-set. The best Hyper-parameters set is therefore trained for 7 Million steps more (almost 22 hours) on the training data-set to further improve the model and to try to achieve the convergence on the best policy. The training Tensorboard data are shown in Figure 5.10.

The charts in Figure 5.10 display policy training results that seem to converge to an optimal policy. The policy optimization is considered satisfying according to general training reading procedures [43], although the approximate Kullback–Leibler divergence (approx_kl) seems to be slightly unstable at the end of the training. High values of Kullback–Leibler divergences can potentially lead to poor policy optimizations. However, the episode reward mean (ep_rew_mean) reaches very high values and seems to be stable, as well as the entropy loss which gets closer to zero and remains stable. The resulting trained model reaches a cumulative reward score of 400000 in the testing data-set, proving that the strategy of increasing the number of training time-steps was favourable to improve policy optimization. It also suggests that higher training time-steps numbers could probably further improve the model. The supply/demand load performances reached favourable

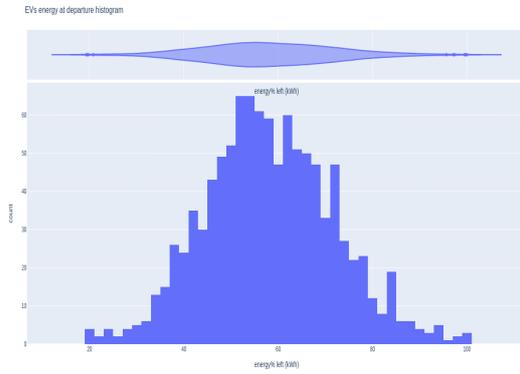
(a) Random actions

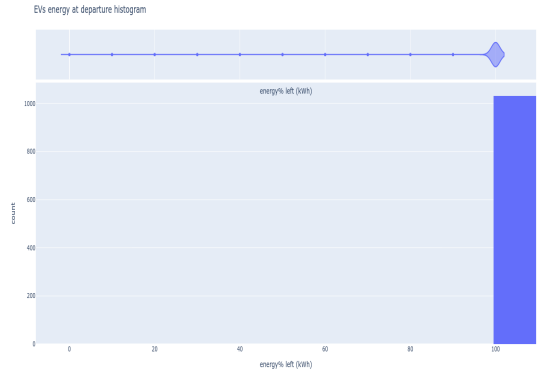
(b) Elvis

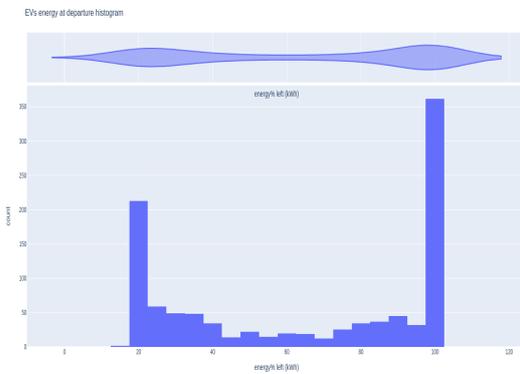
(c) A2C

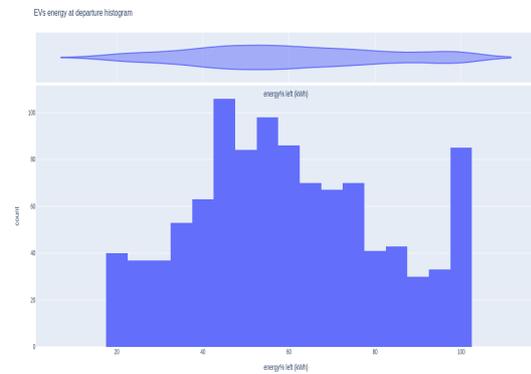
(d) MaskablePPO

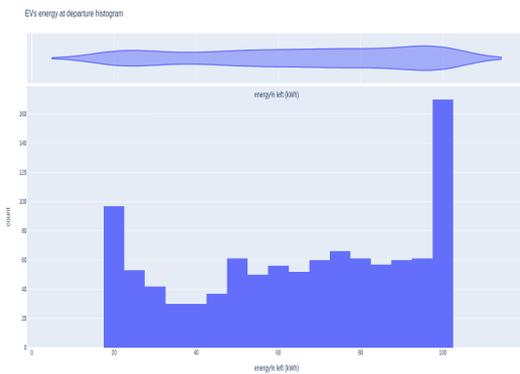
(e) TRPO

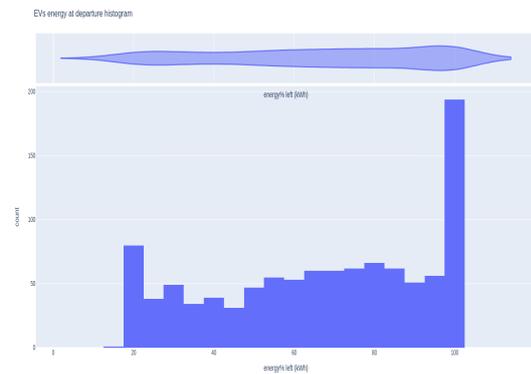
(f) RecurrentPPO

Figure 5.8: EVs energy at departure histograms, algorithms comparison.

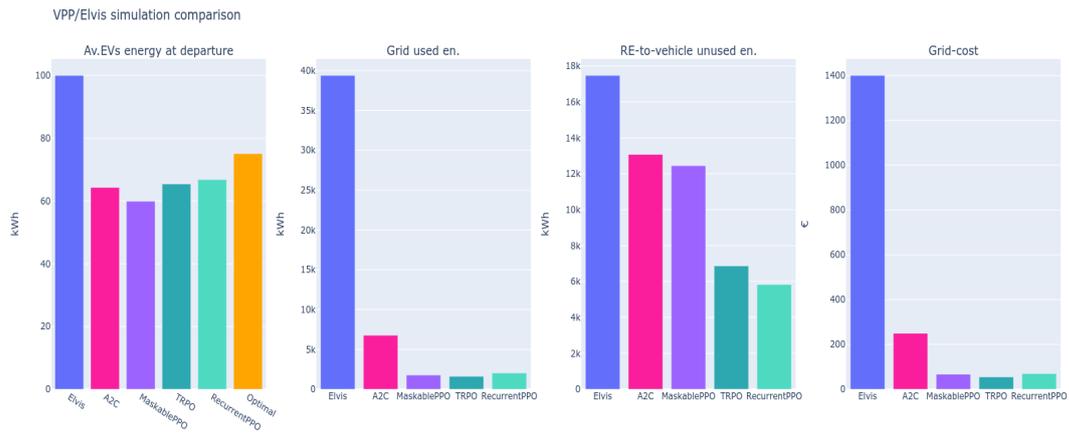

(a) Tuned algorithms results comparison bar-chart

| Algorithm/param. | RE-to-vehicle unused en. | Grid used en. | Grid-cost | Av.EVs energy at departure | Cumulative reward |
|---|---|---|---|---|---|
| Elvis | 17475.87 | 39397.31 | 1400.65 | 100 | 0 |
| A2C | 13078.36 | 6757.376 | 249.43 | 64.31 | 153080.9 |
| MaskablePPO | 12445.26 | 1769.706 | 66.52 | 59.92 | 227422.6 |
| TRPO | 6867.678 | 1601.168 | 54.79 | 65.4 | 357075.1 |
| RecurrentPPO | 5828.97 | 2035.55 | 69.392 | 66.8 | 384691.27 |
| Optimal | 0.01 | 0.01 | 0.01 | 75.08 | 450000 |

(b) Tuned algorithms results comparison table

Figure 5.9: Algorithms trained with best Hyper-parameters set, results comparison

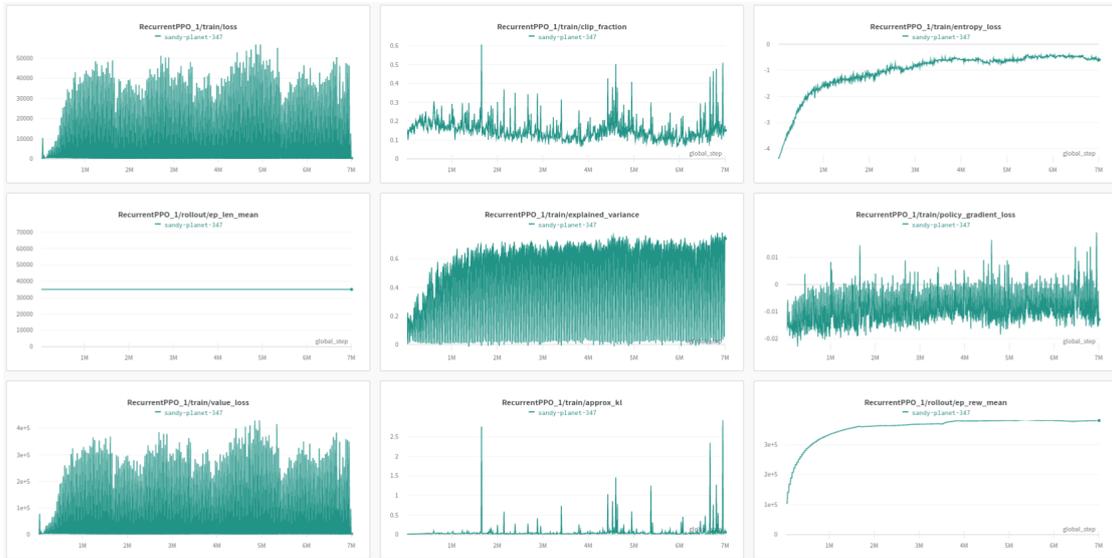

Figure 5.10: RecurrentPPO 7 Million steps training tensorboard

results, as shown in the peak load occurrences comparison with the Elvis simulation in Figure 5.11.

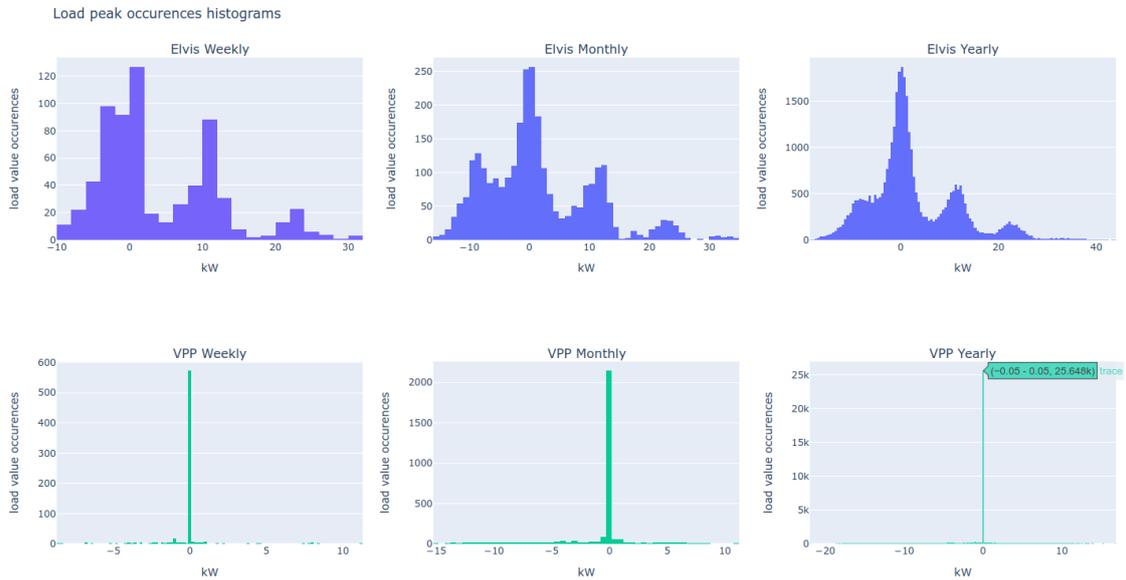

Figure 5.11: 7M trained RecurrentPPO load occurrences comparison with Elvis simulation

The image shows that, for the yearly load distribution, the histogram counts 25648 out of 35041 time-steps (73.1% of the simulation) with a balanced supply/demand load (comprised between -0.05 and 0.05 kW). A detailed observation of the load distribution is described in the chapter on the VPP parameters tuning. Figure 5.12 displays the plot over time of the supply/demand loads that balance each other out thanks to the EVs' charging/discharging loads. Although the negative and positive load profiles (respectively, supply and demand profiles) are not exactly symmetrical, they are generally compensating each other.

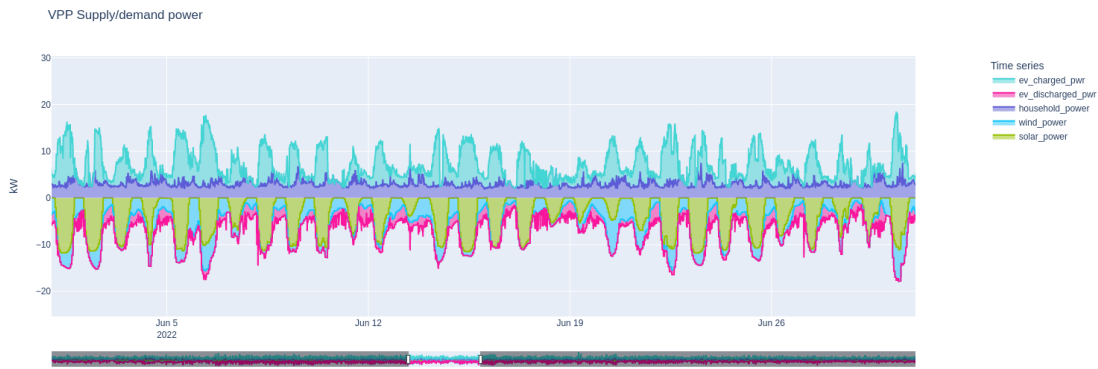

Figure 5.12: RecurrentPPO supply/demand sources results over time.

# Chapter 6

# Model training results

This chapter describes and explains the results obtained by the trained models. It aims to present some benchmarks that can support the algorithm choices and evaluate the general performance of the Virtual Power Plant solution proposed in the Methodology section.

## 6.1 Algorithms comparison

One of the most effective ways to visualize the algorithms results is to plot them in the three-dimensional space of the three main simulation output parameters: the Grid energy used (grid import or "overconsume"), the RE-to-vehicle (RE2V) unused energy (grid export or "underconsume") and the average EVs available energy at departure. the Grid energy cost is highly correlated to the Grid energy used, and therefore it is not a meaningful axis. To compare their performance, each Hyper-parameter tuning simulation outcome is plotted in such space, highlighting the best outcomes for each algorithm and the maximum average EVs available energy point (described in the analysis chapter, for the testing data-set: 0.01 kWh, 0.01 kWh, 75.08 kWh). The presentation of such a scatter plot dividing the algorithms' performance by color is shown in Figure 6.1. The plotted interpolating plane serves purely to help visualize in three dimensions the scattered points.

The distance from the optimal average EVs available energy point is the key to reading the plot. The A2C cluster of results is the most distant from it, as it was the algorithm performing worst. The RecurrentPPO and TRPO clusters are instead getting closer to the optimal point, with their best-trained models being the closest ones. Figure 6.2 presents the same scattered points, but with the size of the points proportional to the cumulative reward obtained in the simulation.

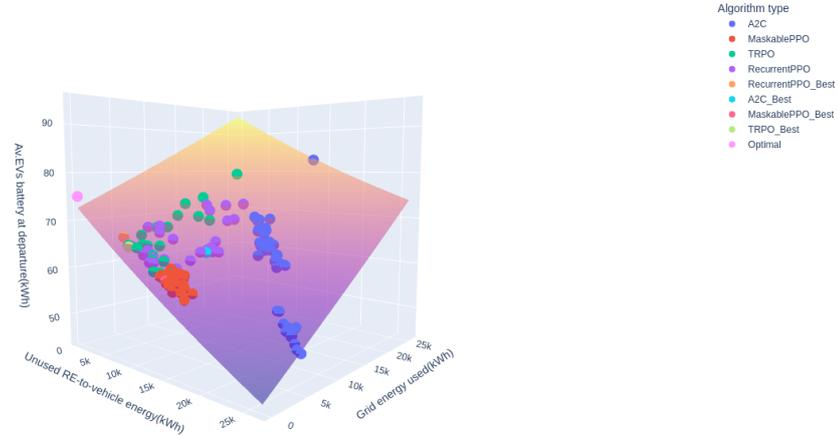

Figure 6.1: Algorithms results 3D bubble plot.

This plot helps to visualize how the cumulative reward increases when getting closer to the origin of the Grid-energy-used axis and the Unused-RE2V-energy axis. This obviously corresponds to the simulation results getting closer to zero for those parameters. Then, as the average EVs available energy at departure increases (going up on the z-axis), the cumulative reward grows even more.

This behaviour is highlighted in the plot of Figure 6.3, where the scattered points' color shade is proportional to their cumulative reward. The plotted results get brighter as they get close to the optimal point, meaning that the reward system was correctly set. However, there is a considerable gap between the best performance from the RecurrentPPO VPP-model and the optimal, even if the optimal point is thought to be unreachable due to the simulation's limits as explained in the analysis chapter.

This distance is highlighted by observing the results in the 2D side plane of the Unused RE2V energy and the average EVs available energy at departure axes, as shown in Figure 6.4. Even without the information on the Grid energy used axis from this side view of the plot, it is possible to correlate the distance from the optimal point with a decreasing cumulative reward, as shown in Figure 6.5.

An interesting pattern can be noticed by plotting the results on the same plane with the color shade being proportional to the Grid energy used, and the size of the scattered points being proportional to the cumulative reward as shown in Figure 6.6. The plot reveals that as the VPP simulations get similar Grid energy used values, the results align in a regression line that confirms the negative proportionality between the Unused RE2V energy and the average EVs available energy at departure axes. This negative proportionality was already noticed in the algorithms tuning results of the previous chapter. The features highlighted in the comparison of the results help visualize the objectives' priority of the VPP agent:

- First, minimizing the Grid energy used to get in the minimum Grid energy used regression line as shown in the plot of Figure 6.6.

- Secondly, minimizing the Unused RE2V energy to exploit more energy and maximizing the average EVs available energy at departure.

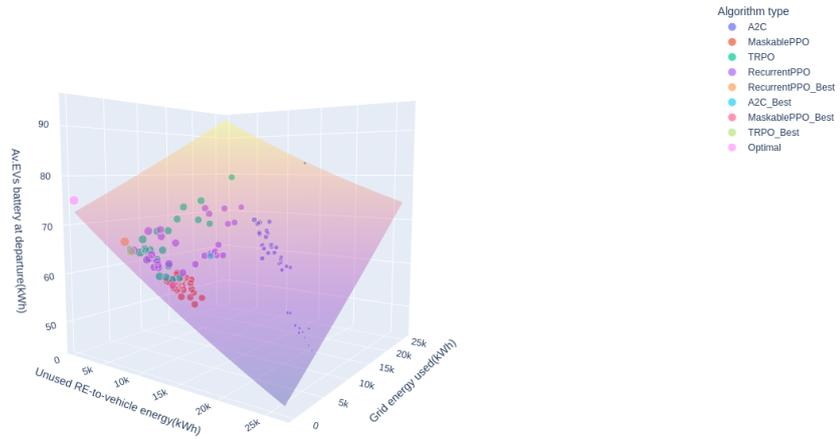

Figure 6.2: Algorithms results 3D plot with size being proportional to their cumulative reward.

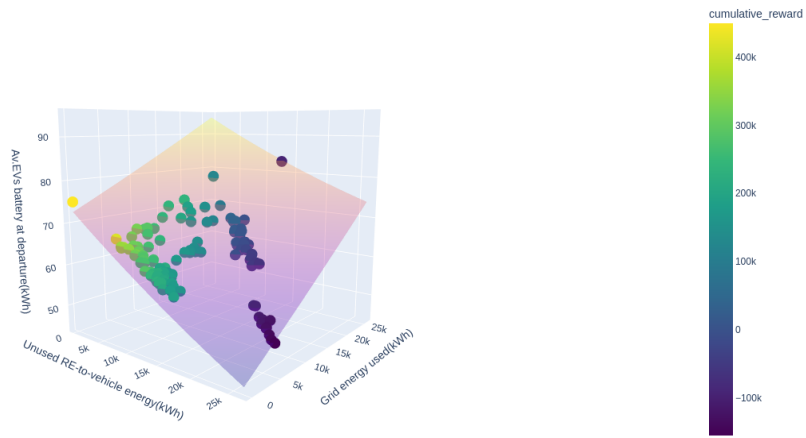

Figure 6.3: Algorithms results 3D plot, with color scale mapped to cumulative reward.

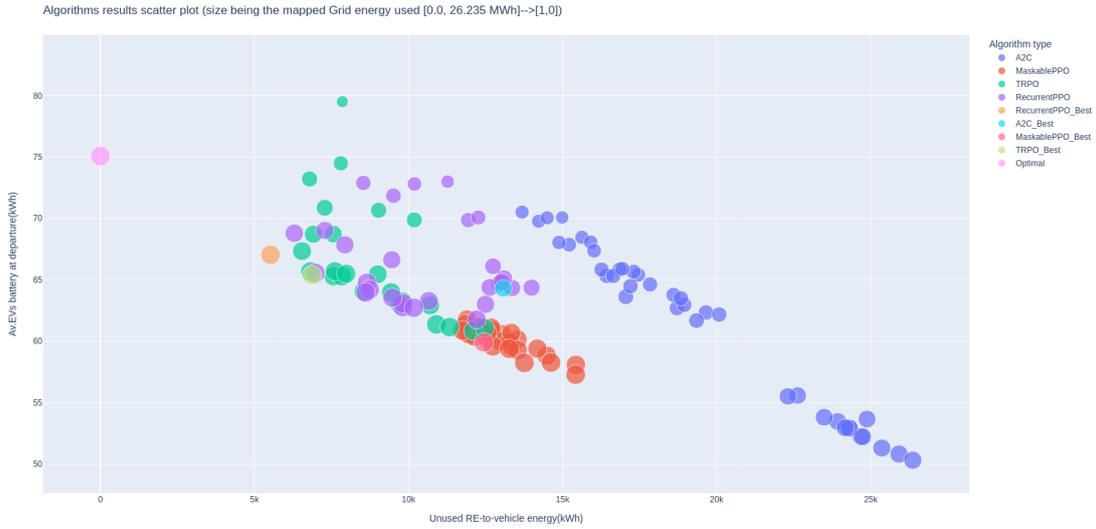

Figure 6.4: Algorithms results 2D plot.

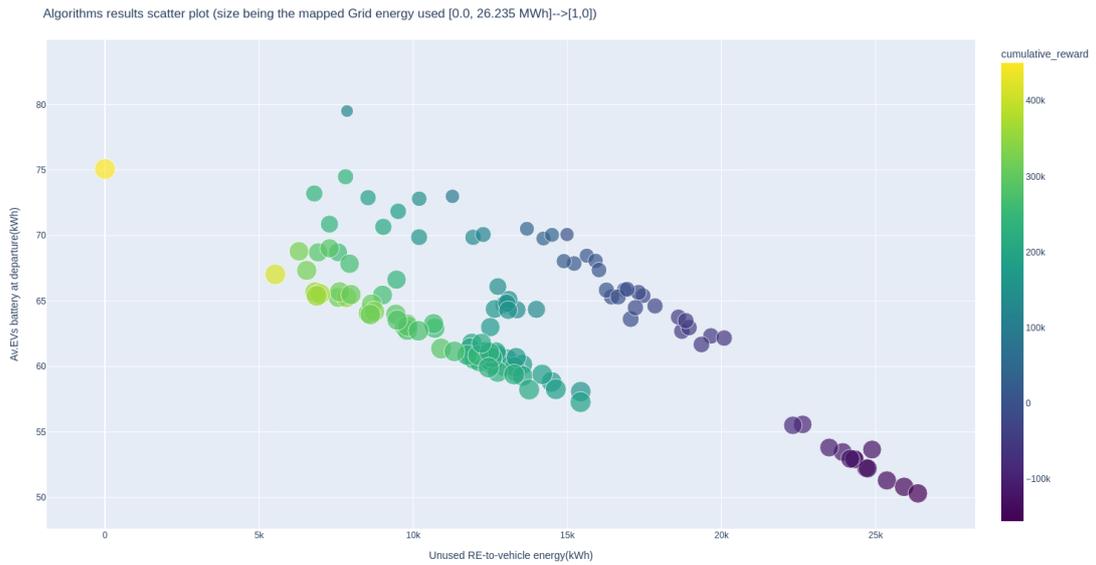

Figure 6.5: Algorithms results 2D plot with cumulative reward color scale.

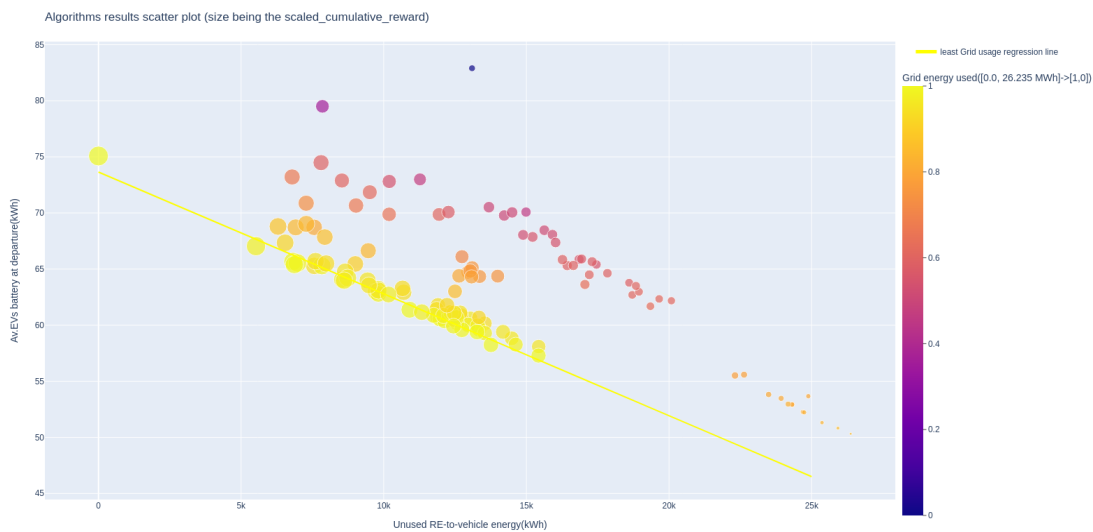

Figure 6.6: Algorithms results 2D plot with Grid energy used color scale.

## 6.2 Environment limits

The scattered plots presented in this chapter demonstrate how the VPP simulation results move in the proposed space according to precise directions. That is why the interpolating plane added in the 3D plot for visualization reasons, actually highlighted the plane where the VPP results lie. Furthermore, the simulation environment limits make it very hard for the VPP agent to get very close to the optimal point. This revelation tells that the Virtual Power Plant has hard limits in terms of minimization of energy waste while also powering EVs when connected to the charging stations. However, this set of VPP agent results is limited to this precise VPP simulation configuration, and it may significantly vary for different ones.

# Chapter 7

# Virtual power plant tuning experiments

This chapter proposes experiments to test the solution proposed for the VPP environment and to tune the Elvis configuration parameters for the best results. It focuses on the EVs' weekly number of arrivals and the effects of different numbers of charging events on the VPP simulation. It concludes with the proposal of the optimal number of weekly EV charging events.

## 7.1 Tuning the virtual power plant

The Elvis EVs infrastructure simulator needs to be set up by loading the configuration builder YAML file (the Elvis configuration file described in the analysis chapter). This file contains the simulation parameters that can be modified to observe different simulation behaviours. There are several parameters to be set, but most of them also modify the VPP environment boundaries initialization and the VPP agent model definition. Nevertheless, there are parameters that can be modified without affecting the VPP environment compatibility:

- The weekly number of EVs arrival or charging events (num_charging_events). It has been set to 20 for the VPP model training and testing.

- The mean parking time, in hours (mean_park). It has been set 23.99, being 24 the maximum limit, since smaller values would create EVs' charging events lasting not enough for sufficient results for a VPP simulation.

- The standard deviation parking time, in hours (std_deviation_park). It has been set to 1 since higher values would make EVs' charging events duration too random for sufficient results for a VPP simulation.

- The mean SoC of EVs at arrival, expressed from 0 to 1, 1 being 100% of the EVs charge capacity (mean_soc). It has been set to 0.5 to simulate realistic EVs' available energies.

- The standard deviation of EVs' arrival SoC (std_deviation_soc). It has been set to 0.1 since higher values would produce EVs' available energies at arrival too random for a realistic simulation.

Among these parameters only the weekly number of EVs' arrivals can be changed without compromising a realistic VPP simulation. Plus, for a fair comparison with the other results, changing only one parameter per experiment is the safest method for meaningful results. Therefore, the following configurations are used to test the VPP trained agent:

- 10 EVs arrivals per week with 50% of average charge.
- 15 EVs arrivals per week with 50% of average charge.
- 20 EVs arrivals per week with 50% of average charge.
- 25 EVs arrivals per week with 50% of average charge.
- 30 EVs arrivals per week with 50% of average charge.
- 35 EVs arrivals per week with 50% of average charge.

The other settings are left unchanged from the previous simulations for every algorithm. However, to diversify the source of VPP simulation data, the simulations of this tuning experiment will be run on the **validating data-set**. The parameters used to compare the results are also the same as seen in the previous chapters. Each of the configurations presented will be used to run 10 VPP simulations. Subsequently, the results for each parameter of each simulation are averaged to obtain an average point per configuration.

## 7.2 Experiments outcome

The simulation outcomes obtained for each configuration are plotted in the 3D parameter space used before, as shown in Figure 7.1. The results are clearly divided into clusters highlighted by the colors of the scattered points, and for each cluster, the average point is plotted with a different color.

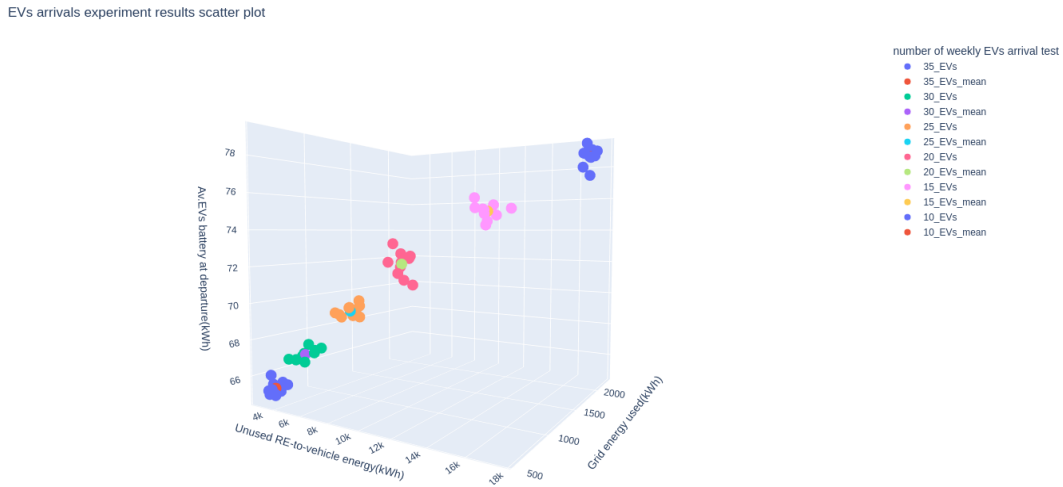

Figure 7.1: VPP tuning experiment results 3D bubble plot.

The alignment of the results suggests a linear correlation between the number of EV charging events per simulation and the position in the parameters' result space. In particular, the plot reveals that a different number of EVs' charging events shifts the simulation outcomes in a line that, for less EVs arrivals, it increases the Grid energy used (grid import), the Average EVs available energy at departure, and the Unused RE2Vehicle energy (grid export). That is, the results get farther away from the objective of minimizing the grid import/export of energy, but the EVs can store more energy since there are fewer of them. However, since the first objective of the VPP is to minimize the grid import of energy, and the second is to minimize the grid export, the cumulative reward obtained for higher numbers of EVs' arrivals is less. This can be seen in Figure 7.2, where the size of the scattered results is proportional to the obtained cumulative reward.

On the contrary, for higher numbers of EVs arrivals, the Grid energy used (grid import), the Average EVs available energy at departure, and the Unused RE2Vehicle energy (grid export) decrease. This leads to generally higher cumulative rewards because the grid import/export decrease. This happens because there are more EVs available throughout the simulation that can provide support to the grid, and the VPP agent has higher chance of doing the right moves. The cumulative rewards of each result can be better observed in the 3D plot of Figure 7.3, where the cumulative reward of each scattered point is mapped into a color scale visible on the right side. As a general trend, higher numbers of EVs' arrivals produce lighter colors for the scattered points.

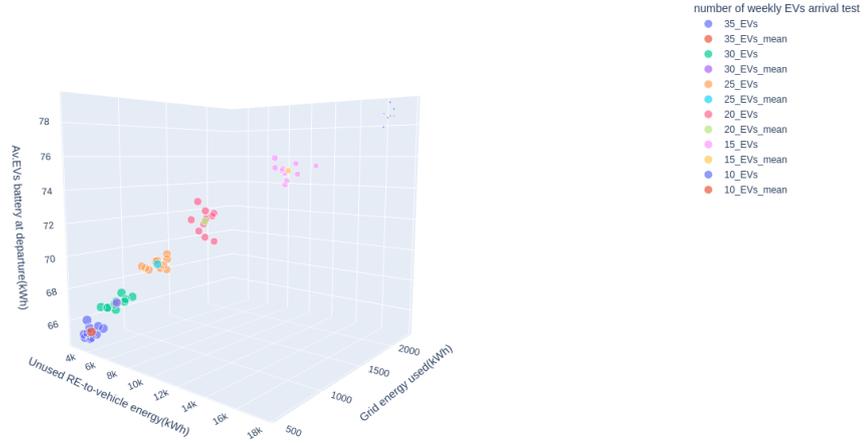

Figure 7.2: VPP tuning experiment results 3D plot with bubble size being proportional to cumulative reward.

However, more EVs arrivals mean that the average amount of energy that EVs can receive from renewable sources is less since the energy produced is limited by the data-set constraints. Therefore, without importing energy from the grid, the Average EVs battery level at departure can only decrease for a higher number of EVs arrival. The linearity of such a decrease can be seen in Figure 7.4, where the results are plotted in the 2D space of Unused RE2vehicle energy (grid export) and Average EVs available energy at departure.

Figure 7.4 describes better the decreasing trend along the axis: the difference between the coordinates of the clusters seems to decrease as the number of EVs arrivals rises. In other words, the clusters of results for 35 and 30 EVs arrivals per week are much closer than the ones for 15 and 10 EVs weekly arrivals. This behaviour is confirmed in Figure 7.5, where the color shades of the scattered points, being mapped into the Grid the energy used (grid import), quickly change in re-sults for 10, 15, and 20 weekly EVs arrivals, while for 25, 30, and 35 it changes slowly.

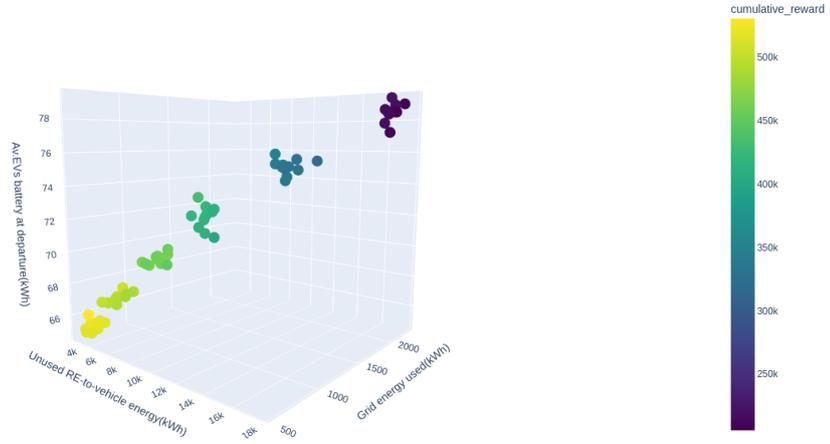

Figure 7.3: VPP tuning experiment results 3D plot with bubble color scale mapped to cumulative reward.

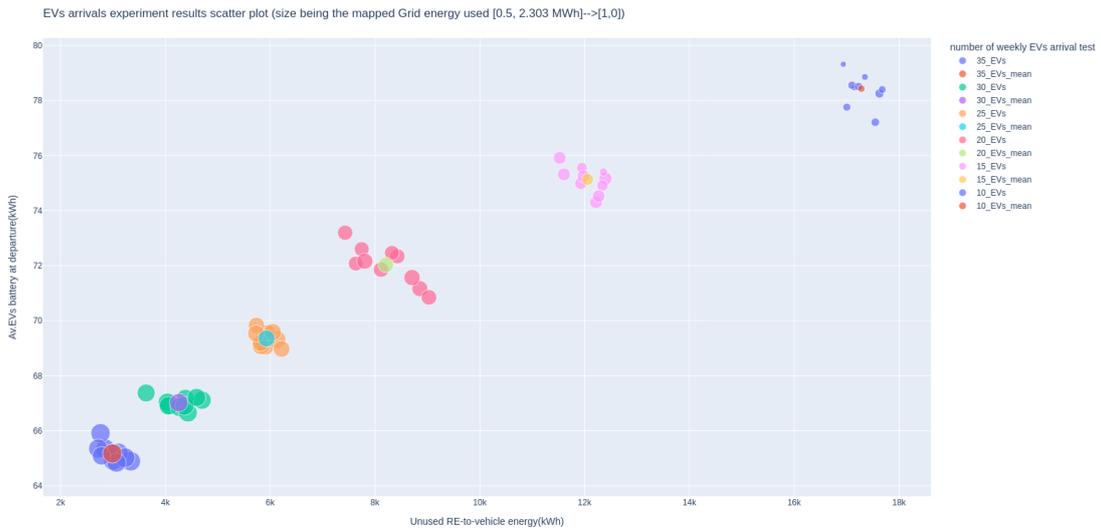

Figure 7.4: VPP tuning experiment results 2D bubble plot.

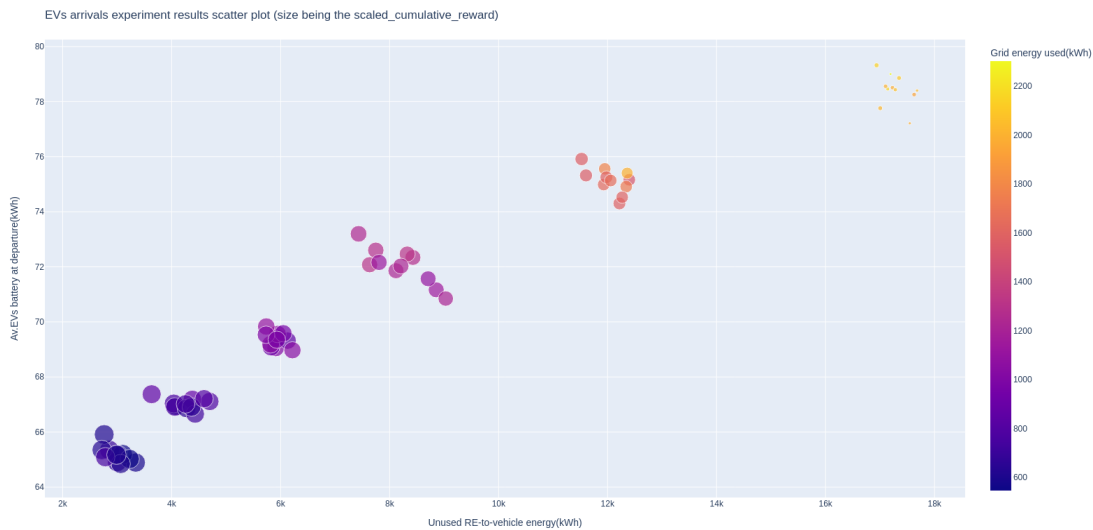

Figure 7.5: VPP tuning experiment results 2D bubble plot with color scale mapped to Grid energy used (grid import).

Figure 7.5 presents the size of the scattered results points proportional to the cumulative reward. In that plot, the size of the clustered results is increasing as the number of weekly EVs' arrivals increases, as it was also seen in Figure 7.3. More specifically, the size of the scattered points rapidly increases for 10, 15 and 20 weekly EVs' arrivals, while it imperceptibly increases between the 25, 30 and 35 weekly EVs arrivals clusters. This different rate of results variation per number of weekly EVs arrival is taken into account for choosing the best configuration settings for the VPP simulation. Nevertheless, another important analysis of the VPP tuning regards the distribution of the EVs' available energies at departure. The results of such distributions are presented in the histograms available in Figure 7.6. The distribution of the EVs' energies available at departure is important to understand the VPP performances on guaranteeing EVs a minimum level of energy. Visualizing such distribution is important since unfortunately, the Average EVs available energy at departure parameter is not descriptive enough for such a matter.

A first look at the histograms of Figure 7.6 suggests that as the number of weekly EVs arrival increases, the lower fence, the median, and the upper fence decrease creating a homogeneous distribution. This can be seen in the marginal distribution displayed on top of each histogram of Figure 7.6: the distribution of the first graph (Figure 7.6.a) is concentrated in the higher EVs energy at departure values, while the last distribution (Figure 7.6.a) has a higher concentration in the intermediate values. In general, the percentage of EVs leaving with high levels of battery energy is higher for the simulations with fewer EVs weekly arrivals. Thus, the distributions' shape confirms that for higher numbers of EVs weekly arrivals the VPP's ability to guarantee decent levels of energy in the EVs gets poor. The histograms' results can be summarized in the table displayed in Figure 7.7.

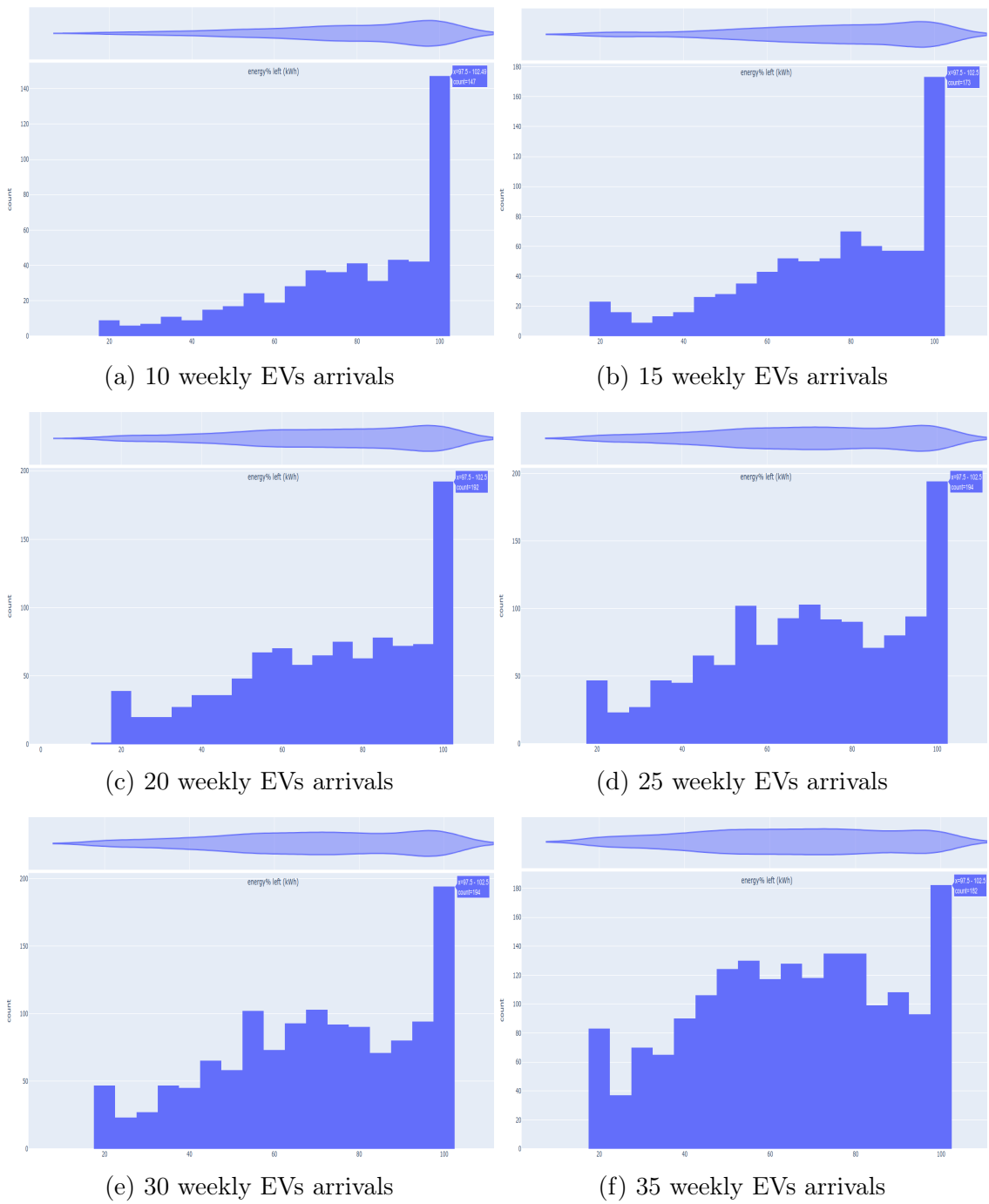

Figure 7.6: EVs energy at departure histograms for different weekly arrival values

The last data to analyze for each configuration is the distribution of load values during the simulation. The distributions of the yearly load values for each weekly arrivals parameter are available in the comparisons displayed in Figure 7.8. In this regard, the resulting load values between -0.1 kW and 0.1 kW are considered very low and grouped together as supply/demand compensated load. The results for the compensated load time-steps are highlighted for each configuration, ranging from 17000 time-steps (48%) for 10 weekly EVs arrivals, up to 30000 time-steps (86%) for 35 weekly EVs arrivals. The remarkable increase of time-steps with supply/demand load balance clearly shows how the number of charging events in the simulation directly affects the performed support to the load.

| Energies at departure table | | Distribution data | | | [EVs battery level at departure] | | | | | | | |
|---|---|---|---|---|---|---|---|---|---|---|---|---|
| Weekly EVs arrival | Tot EVs n. | Q1 (kWh) | Median (kWh) | Q3 (kWh) | battery <22.5% | | 22.5%<battery<52.5% | | 52.5%<battery<97.5% | | Battery≈100% | |
| | | | | | EVs n. | n.EVs% | EVs n. | n.EVs% | EVs n. | n.EVs% | EVs n. | n.EVs% |
| 10 | 522 | 66.98 | 85.64 | 99.74 | 9 | 1.72% | 65 | 12.45% | 301 | 57.66% | 147 | 28.16% |
| 15 | 780 | 60.06 | 79.21 | 96.4 | 23 | 2.95% | 108 | 13.85% | 476 | 61.03% | 173 | 22.18% |
| 20 | 1040 | 55.83 | 74.68 | 93.49 | 40 | 3.85% | 187 | 17.98% | 621 | 59.71% | 192 | 18.46% |
| 25 | 1293 | 52.32 | 71.1 | 90.79 | 43 | 3.33% | 286 | 22.12% | 767 | 59.32% | 197 | 15.24% |
| 30 | 1562 | 49.73 | 68.95 | 87.23 | 61 | 3.91% | 368 | 23.56% | 932 | 59.67% | 201 | 12.87% |
| 35 | 1820 | 47.01 | 65.97 | 84.26 | 83 | 4.56% | 492 | 27.03% | 1063 | 58.41% | 182 | 10.00% |

Figure 7.7: EVs energy at departure distributions table

### 7.2.1 VPP tuning conclusions

The results of the proposed experiment presented the obligation of a trade-off between minimizing the grid import/export energy and maximizing the average EVs available energy at departure. This trade-off is necessary for choosing the best-suited number of weekly EVs arrival in the VPP simulation. The most reasonable choice is to pick an intermediate result that slightly improves the supply/demand balance load. For this reason, the number of weekly EVs arrival chosen, as the optimal parameter for the Elvis configuration, is 25.

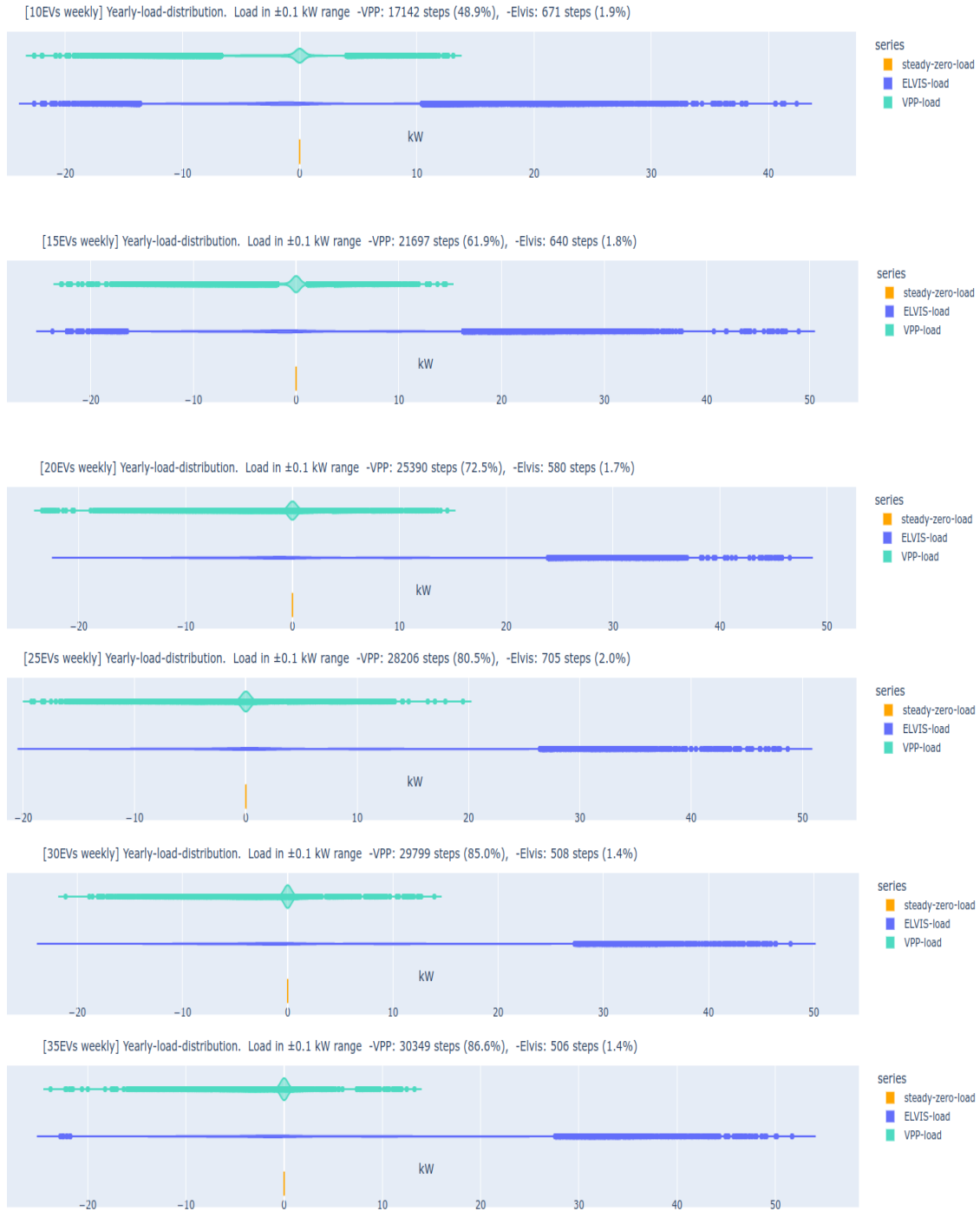

Figure 7.8: VPP load distribution for different weekly arrival values

## 7.3 Final results

The optimal configuration results are now presented to wrap up the optimal trained VPP features and capabilities. As a final evaluation of the VPP, the analysis of the self-consumption and the autarky rates chart is proposed in Figure 7.9. The image presents a comparison of the charts with the corresponding Elvis simulation results, divided in self-consumption over supply energy and autarky over demand energy. The Elvis simulation presents a self-consumption rate of 58.01% and an autarky rate of 43.9% solely with the Renewable-to-households (RE2house) and the Renewable-to-EVs (RE2V) consumption. These rates can be considered already pretty high considering that the simulation does not support Renewable-to-vehicle (RE2V) or Vehicle-to-any (V2X) energy transfers. The VPP simulation instead reaches 90% of self-consumption rate and 98.3% autarky rates considering the RE2house and the RE2V consumption plus the EV-to-EV (V2V) and the EV-to-households (V2B) energy transfers. These rates are considered very high for those parameters, and they are backed up by minor percentages of EV-to-households energy transfer, and significantly high percentages of Renewable-to-EVs transfer concerning the Elvis simulation. However, the EV-to-EV energy transfer shares provide only relative support to the VPP because the energy is shifted from an EV to another one. A bad-looking result is instead shown from the EV-to-grid energy transfer, which means that even if for an amount of energy smaller than 1 MWh, the VPP agent made the mistake of discharging the EVs when the supply was higher than the demand. Overall, the charts reveal good results and the VPP agent managed to minimize the RE-to-grid export to 8.43% of the total supplied energy, and to minimize the Grid-to-house and Grid-to-EV import to 0.4% and 1.3%, respectively.

The VPP directions of loads can be observed in the supply/demand load plot of Figure 7.10. In general, the VPP manages to follow the zero net load by balancing the supply and demand, even if sometimes the RE-to-EV self-consumption does not cover the whole renewable production, as it can be seen by the two cyan lobes with a rift filled with blue-violet (RE-to-grid load). A positive result is the fact that the Grid-to-households and Grid-to-EVs load is almost constantly zero, while the EV-to-EV load with unclarified effects is slightly present from time to time (highlighted in yellow).

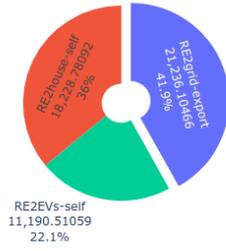
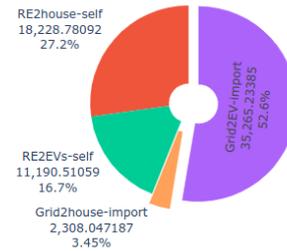
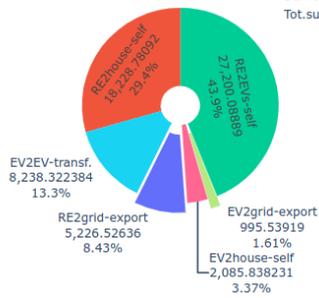
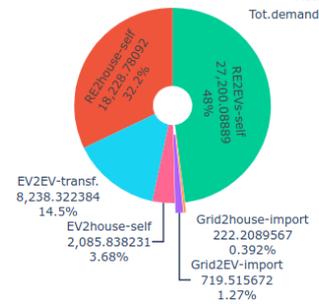

Figure 7.9: VPP self-consumption and autarky charts

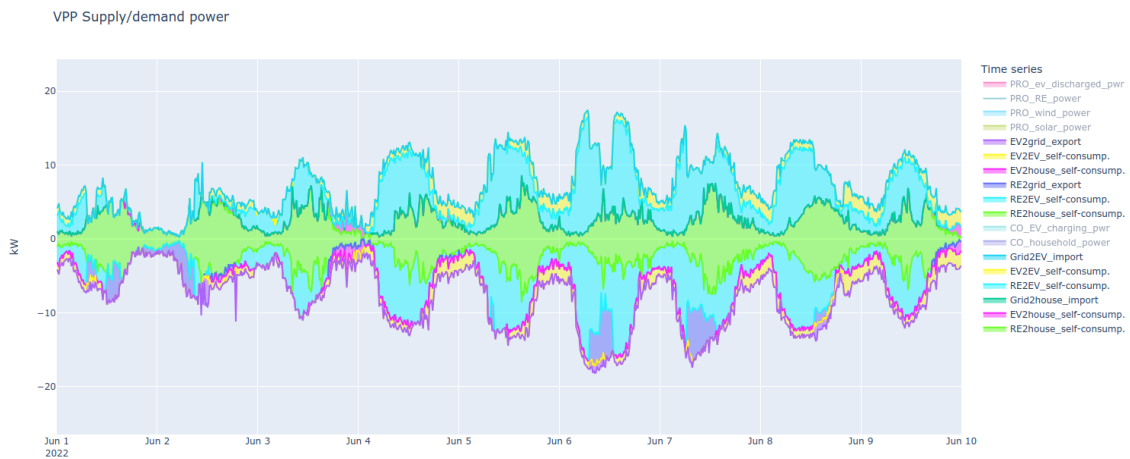

Figure 7.10: VPP supply-demand with directions of energy.

# Chapter 8

# Conclusions and future work

This chapter summarizes the analysis of the obtained results in this project. It formulates a conclusion for the Virtual Power Plant solution implemented and introduces future work perspectives to further evaluate and utilize the findings of this thesis.

## 8.1 Conclusion

The final results presented in the previous section suggest an excellent outcome for this research since they reveal very high percentages of self-consumption and autarky. These positive results are considered a success for the reinforcement learning solution implemented to satisfy the primary goal of balancing the supply/demand. However, the results regarding the energy left in the EVs at departure are satisfying, but not brilliant, because the VPP cannot guarantee high levels of energy left at departure for all the EVs. The trade-off between the supply/demand balance and the energy left on the EVs was clear since the beginning of the research. Now, the limits of the trade-off are clear as well.

Another important conclusion concerns the choice of reinforcement learning as the orchestrating core of the virtual power plant. Reinforcement learning science is actually a branch of optimal control theory as explained in the background chapter. For this reason, the RL agent actions began to make sense and the agent started to learn only when the **input data** and the **reward functions** were rephrased as for an optimal control problem. Specifically, when the reward function and the adaptive charging power were re-shaped to keep the resulting load as close as possible to zero. Any other rewards summed for secondary objectives resulted in poor learning. Nevertheless, even after training and tuning the best algorithm, by analyzing each action performed, the agent would still make mistakes from time to time, because the policy optimization of the algorithm used is still based on probabilities and states exploration. The random component of the action selection of the policy optimization was supposed to fit the aleatory variable of the EVs' presence and available energy. However, balancing the supply/demand load is a problem that can be optimally solved with control theory and a solution based on this would produce better results, despite the EVs' aleatory presence.

## 8.2 Solution limitations

The solution proposed is strictly limited by the data-sets, which are built for the simulation considering a limited amount of energy produced by renewable sources. The renewable energy infrastructure used by this research is composed of real modules available on the market. These modules can be assembled in real life without an elevated cost for the current technologies, considering a shared residential area of 4 households. However, since this solution for smart EVs' charging stations is thought to be applied when the EVs demand will raise in 2030, the future available technologies could provide significantly better performances, and thus the capacity of the simulation could be adjusted on such improvements. In any case, modifying the infrastructure capacity would change the limits of the VPP and would modify the simulation outcomes, particularly for the energy stored in the EVs.

Furthermore, the performances of the VPP are tied to the RL algorithms used, which come from open-source repositories with limited support for the designed environment. For this reason, implementing the algorithms from scratch could potentially lead to better results. For instance, implementing a model with a learning algorithm based on value-function methods, rather than policy gradient methods, would fit more the deterministic type of the VPP agent policy, as explained in the algorithm background theory chapter.

## 8.3 Future Work

This research was focused on the use of EVs as storage devices for renewable energy resources, without implementing any other fixed storage device. This choice was taken to concentrate on the study of EVs' behaviour as storage devices, but new possible research should be conducted on a similar environment that includes storage devices to provide additional support to balance the supply and demand of energy. In a broader view, the VPP solution proposed uses EVs, PV modules and WT modules as the only components of the VPP subject of research, but they can be the base for introducing any sort of distributed energy device (storing/producing).

To conclude, re-implementing the VPP solution based on control theory rather than reinforcement learning could also lead to promising improvements, for the reasons described in the previous section.

# Appendix

## .1 Source code

The source code repository contains the developed Python script of the Virtual Power Plant environment. In the repository, there are the Jupyter notebooks with training, tuning, experiments and simulations results already loaded for reference. A detailed description of how to open the notebooks on Google Colab and the direct access links to the notebooks are available in the README.md file. The notebooks can also be run again to study different outcomes. The source code repository of this research project is available at:

https://github.com/francescomaldonato/RL_VPP_Thesis